\documentclass[twocolumn]{aastex62}

\usepackage{txfonts}
\usepackage{graphicx}
\usepackage{color}
\usepackage{multirow}
\usepackage{rotating}

\newsavebox{\savefig}

\newcommand{\dn}{DN\,s$^{-1}$\,px$^{-1}$}

\begin{document}

\title{Imaging evidence for solar wind outflows originating from a CME footpoint}

\correspondingauthor{J. L\"{o}rin\v{c}\'{i}k}
\email{lorincik@asu.cas.cz}

\author[0000-0002-9690-8456]{Juraj L\"{o}rin\v{c}\'{i}k}
\affil{Astronomical Institute of the Czech Academy of Sciences, Fri\v{c}ova 298, 251 65 Ond\v{r}ejov, Czech Republic}
\affil{Institute of Astronomy, Charles University, V Hole\v{s}ovi\v{c}k\'{a}ch 2, CZ-18000 Prague 8, Czech Republic}

\author[0000-0003-1308-7427]{Jaroslav Dud\'{i}k}
\affil{Astronomical Institute of the Czech Academy of Sciences, Fri\v{c}ova 298, 251 65 Ond\v{r}ejov, Czech Republic}

\author[0000-0001-5810-1566]{Guillaume Aulanier}
\affil{LESIA, Observatoire de Paris, Universit\'e PSL , CNRS, Sorbonne Universit\'e, Universit\'e de Paris, 5 place Jules Janssen, 92190 Meudon, France}

\author[0000-0003-3364-9183]{Brigitte Schmieder}
\affil{LESIA, Observatoire de Paris, Universit\'e PSL , CNRS, Sorbonne Universit\'e, Universit\'e de Paris, 5 place Jules Janssen, 92190 Meudon, France}
\affil{Centre for Mathematical Plasma Astrophysics, Dept. of Mathematics, KU Leuven, 3001 Leuven,
Belgium}
\affil{SUPA School of Physics and Astronomy, University of Glasgow, Glasgow G12 8QQ, UK}

\author[0000-0001-9638-3082]{Leon Golub}
\affil{Smithsonian Astrophysical Observatory, 60 Garden Street, Cambridge MA, 02138, USA}

\begin{abstract} 
We report on the Atmospheric Imaging Assembly (AIA) observations of plasma outflows originating in a coronal dimming during the 2015 April 28th filament eruption. After the filament started to erupt, two flare ribbons formed, one of which had a well-visible hook enclosing a core (twin) dimming region. Along multiple funnels located in this dimming, a motion of plasma directed outwards started to be visible in the 171\,\AA~and 193\,\AA~filter channels of the instrument. In time-distance diagrams, this motion generated a strip-like pattern, which lasted for more than five hours and which characteristics did not change along the funnel. We therefore suggest the motion to be a signature of outflows corresponding to velocities ranging between $\approx70$ and 140 km\,s$^{-1}$. Interestingly, the pattern of the outflows as well as their velocities were found to be similar to those we observed in a neighboring ordinary coronal hole. Therefore, the outflows were most likely a signature of a CME-{induced solar} wind flowing along the open-field structures rooted in the dimming region. Further, the evolution of the hook encircling the dimming region was examined in the context of the latest predictions imposed for the three-dimensional magnetic reconnection. The observations indicate that the filament's footpoints were, during their transformation to the dimming region, reconnecting with surrounding canopies. To our knowledge, our observations present the first imaging evidence for outflows of plasma from a dimming region.
\end{abstract}


\section{Introduction} \label{sec_intro}

Solar flares are sudden releases of magnetic energy in the solar atmosphere. They can be accompanied by eruptions of coronal material into the interplanetary space called the Coronal Mass Ejections \citep[CMEs; see e.g.,][]{cliver95, fletcher11}. CMEs can be associated with eruptions of filaments, or prominences when observed above the limb \citep[e.g.,][]{schmieder02}. After the erupting material escapes into the interplanetary space, volumes devoid of emission are often observed in regions where it used to reside \citep[e.g.,][]{thompson00, howard04, webb12}. Referred to as the coronal dimmings, their name originates in an abrupt decrease of plasma intensity observed in various parts of the electromagnetic spectrum, such as in the soft X-rays \citep[e.g.,][]{hudson96,sterling97} or EUV \citep[e.g.,][]{thompson98,zarro99}. 

Even though the dimmings were originally thought to be caused by drops of the temperature, the deficit of emission is mainly caused by the depletion of plasma along expanding field lines \citep{harrison00, harra01, harrison03}. Mass lost from dimming regions reaches up to 70\% of that contained in the CME \citep[e.g.,][]{howard04, jin09}. From a morphological point of view, the dimming regions resemble coronal holes. They are however relatively-smaller and have a shorter lifespan, for which they are sometimes referred to {as transient} coronal holes \citep[TCHs;][]{rust83, kahler01}.

The literature distinguishes between core and secondary dimmings \citep[see e.g.,][]{dissauer18b}. Core dimmings can be found along the flaring polarity inversion line (PIL). They are typically observed {in pairs, and are therefore usually referred} to as twin dimmings. They correspond to expanding legs of erupting CMEs \citep{hudson96, gopalswamy00}. Secondary dimmings can be located further away from PILs, but they are still associated with the CMEs as they are believed to be footpoints of loops dragged by the erupting material \citep{chertok05, mandrini07, dissauer18a}. 

Supporting evidence for the mechanism causing the formation of the dimming regions was confirmed by spectroscopic observations, which revealed the dimmings to be sources of blueshifted spectra. They were first evidenced by \citet{harra01}, who found blueshifts of 30 and \mbox{100 km\,s$^{-1}$} for lines formed at coronal and transition region temperatures, respectively. Ever since, numerous studies reported on outflows (upflows) in either core or secondary dimmings, with velocities of the order of tens \citep[e.g.,][]{harra07, imada07, jin09, attrill10, miklenic11} or even a hundred km\,s$^{-1}$ \citep[e.g.,][]{imada07, tian12, veronig19}. Variations in outflow velocities throughout the dimming regions are apparent in most of the rasters produced by \textit{Hinode}/EIS. Sources of the strongest outflows typically correspond to plages \citep{imada07} or coronal loop footpoints \citep{harra07, jin09, attrill10}. 

The outflows from dimming regions have usually been assumed to be a signature of the CME-driven expansion, stretching, or opening of magnetic fields \citep[see e.g.,][and references therein]{veronig19}. According to \citet[][]{cheng16} and \citet[][]{vanninathan18}, most of the dimming's material is depleted within tens of minutes after the CME's onset {and in} some of the spectroscopic studies, the immediate causality between the CME and the outflows {is not} apparent. For example, \citet{harra07} observed {weak outflows existing prior to the analysed flare along structures neighboring with the studied active region}. Outflows following the CME analysed by \citet{attrill10} were observed for {at least nine hours after the eruption (Figure 9 therein). \citet{mcintosh10} report on outflows developing along opening field lines rooted in the dimming region associated with the CME. Authors however mention that the outflows, though weaker, might have been present along loops rooted therein even before the eruption}. Last but not least, TCHs have been known as one of the sources of the solar wind \citep{kahler01, jana08} since the Yohkoh era. The outflows observed in the dimming regions might therefore be signatures of process other than the CME-driven expansion of the magnetic fields.

According to \citet{jin09}, the outflows develop to refill the dimming region with transition-region (TR) plasma. The scenario proposed by \citet{mcintosh10} describes the outflows as {the secondary source of the mass loaded into the erupting CME. They speculate that the outflows can be signatures of fast wind developing behind the CME along spicules, where the Alfv\'en wave energy dissipates.} \citet{harra07} and \citet{attrill10} supposed that the long-lasting outflows from dimming regions might be signatures of the solar wind. Besides, \citet{attrill10} noted that outflows contributing to the solar wind {had already} been observed at boundaries between active regions and coronal holes \citep[e.g.,][]{baker07, sakao07, harra08}, which was confirmed by \textit{in-situ} measurements \citep[e.g.,][]{kojima99, ko06, culhane14, kilpua16}. At the boundaries between the open- and closed-field line structures, the so-called interchange reconnection \citep{fisk01, crooker02, fisk05, fisk06, delzanna11} acts in further opening of field lines composing the closed-line structures. \citet{attrill08} suggested that the interchange reconnection should also occur between the open-field lines of dimmings and small coronal loops. Even though this processes was not related to the dimming region outflows, the authors {proposed} its role in recovering of dimming regions. CME-driven interchange reconnection was also analysed by \citet{zhu14, yang15}, and \citet{zheng17}. The similarity between the characteristics of the outflows observed in dimmings and at the edges of active regions was also discussed by \citet{tian12}. The authors mention that both phenomena might be driven by a similar mechanism. For example, magnetohydrodynamic (MHD) simulations revealed that outflows with velocities comparable to the observed ones can be produced as a consequence of magnetic reconnection \citep{roussev01, ding11}. 

A deeper understanding of mechanisms causing the dimming region outflows might be brought in by their continuous imaging observations, exploiting higher temporal and spatial resolution compared to spectroscopic ones. Even though the imaging observations were often used to provide context for the spectroscopic data, outflows have not yet been found using imaging data alone. It is worth questioning if the outflows within the dimming regions could be observed simultaneously with the evolution of the structures within and in the vicinity of the dimming region. 

Core dimming {regions are located} near the ends of solar flare ribbons \citep[e.g.,][]{miklenic11}, where they are encircled by ribbon hooks \citep[e.g.,][]{dudik14}. Since flare ribbons map reconnection activity \citep[e.g.,][]{fletcher04, lee08, qiu17, janvier17, lorincik19a} their investigation provides information, among other, about the role of the magnetic reconnection in the evolution of dimmings. 

Recent insights into the physics of flare ribbons were mainly provided by 3D MHD simulations of solar eruptions \citep[e.g.,][]{schrijver11, savcheva12, inoue15}. An example of a model capable of reproducing hooked ribbons is the standard flare model in three dimensions \citep{aulanier12,aulanier13,janvier13}. It describes evolution and eruption of an unstable magnetic flux rope \citep[see the review of][and references therein]{janvier15}. This model has recently been used to investigate the geometries in which field lines reconnect with respect to the actual positions of ribbons and hooks \citep{aulanier19}. {Among other reconnection geometries, the} authors introduced the so-called \textit{ar--rf} reconnection {($a$=arcade, $r$=flux rope, $f$=flare loop, Section \ref{sec_arrf_mech}}) between field lines composing and overlying the erupting flux rope, both rooted near the hook. It has been evidenced \citep[][]{zemanova19, lorincik19b, chen19} that this reconnection results in the expansion of the hooks and therefore the areas of the dimmings. Indications for the reconnection occurring in this geometry have however never been observed simultaneously with the dimming region outflows.

In this manuscript, we present the first imaging evidence for outflows from a core dimming region. We analyse their characteristics and discuss which mechanisms might lead to their existence. We focus on both the short- and long-term evolution of the hook, dimming region within, and surrounding corona, in the context of the predictions imposed by {the standard flare model in three dimensions}. We also discuss whether the outflows from the dimming region can be associated with the three-dimensional ar--rf reconnection. 

This manuscript is structured as follows. Sections \ref{sec_data} and \ref{sec_observations} briefly describe the data and introduce the analysed eruption. In Section \ref{sec_outflows}, we present the observations of the outflows in the dimming region and {a nearby} coronal hole, while Section \ref{sec_interpretation} {considers} the interpretation of the outflow observations. Section \ref{sec_arrf} is focused on the observations of the hook and the surrounding corona in the context of 3D magnetic reconnection. The summary is presented in Section \ref{sec_conclusions}.

\begin{figure*}[h]
  \centering    
    \includegraphics[width=17.00cm, clip,   viewport=  15 50 470 566]{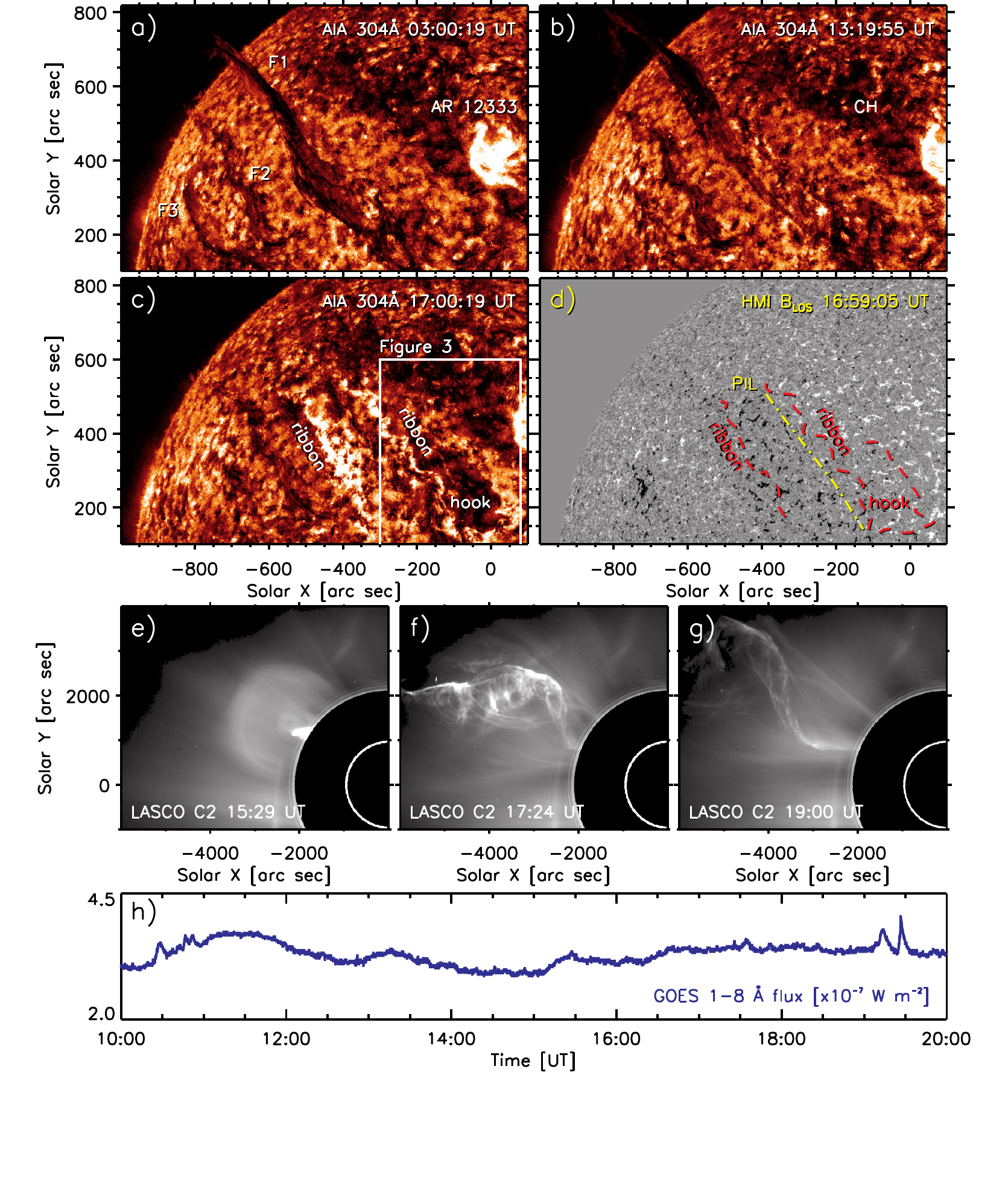}
  \caption{Portion of the solar disk in the 304\,\AA~filter channel data of \textit{SDO}/AIA (panels (a)--(c)) before, during, and after the 2015 April 28th filament eruption. In panel (a), three quiescent filaments F1--F3 and the NOAA 12333 active region are { marked. The} acronym CH marks the neighboring coronal hole. Panel (c) shows the negative- and positive-polarity ribbons and the positive-polarity ribbon hook. The distribution of the line-of-sight component of the underlying magnetic field $B_{\text{LOS}}$ observed using \textit{SDO}/HMI is shown in panel (d). The $B_{\text{LOS}}$ data are saturated to $\pm$30\,G. The yellow dash-dotted line indicates PIL. Panels e--g show three different stages of the eruption viewed through the LASCO C2 coronagraph. Panel (h) shows \textit{GOES} hard X-ray flux measured in the \mbox{1--8\,\AA~} channel. \label{fig_overview} \\ {Animated version of the 304\,\AA~filter channel observations (panels (a)--(c)) is available in the online journal. The animation starts at 03:00 UT, ends at 18:00 UT, and its real-time duration is 15 seconds.} } 
\end{figure*}

\section{Data} \label{sec_data}

We analyse an eruption of a quiescent filament, which took place on 2015 April 28th. To do so, we primarily use data from the Atmospheric Imaging Assembly \citep[AIA,][]{lemen12} onboard the \textit{Solar Dynamics Observatory} (SDO). 

AIA provides full-disk observations of the solar atmosphere in a multitude of filter channels sensitive to the temperatures of the order of 10$^4$ to 10$^7$ K. The solar photosphere is observed in the 1700\,\AA~and 4500\,\AA~filter channels, the chromosphere and the transition region in the 304\,\AA~channel and the 1600\,\AA~filter channels, respectively, and the corona in the 171\,\AA~, 193\,\AA~, 211\,\AA~, and 335\,\AA~filter channels. Finally, data from the 94\,\AA~and 131\,\AA~are typically used to study the hot emission {of solar} flares. The pixel size of AIA is 0.6\arcsec, its spatial resolution is $\approx$1.5\arcsec and the data are typically taken at a cadence of 12\,s or 24\,s, depending on the filter channel. 

\begin{figure*}[t]
  \centering   
    \includegraphics[width=8.00cm, clip,   viewport= 15 0 470 340]{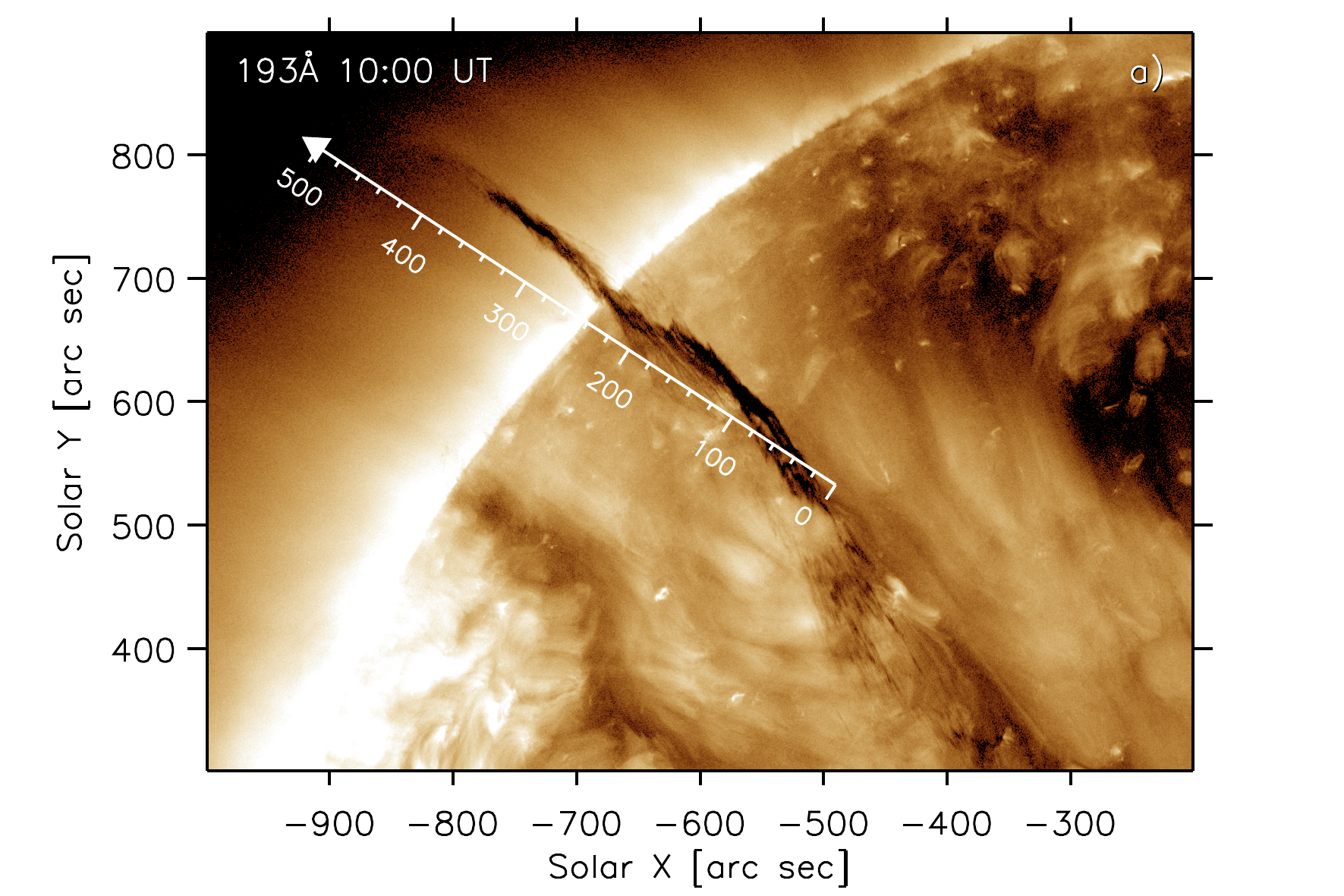} 
    \includegraphics[width=9.00cm, clip,   viewport= 0 0 510 340]{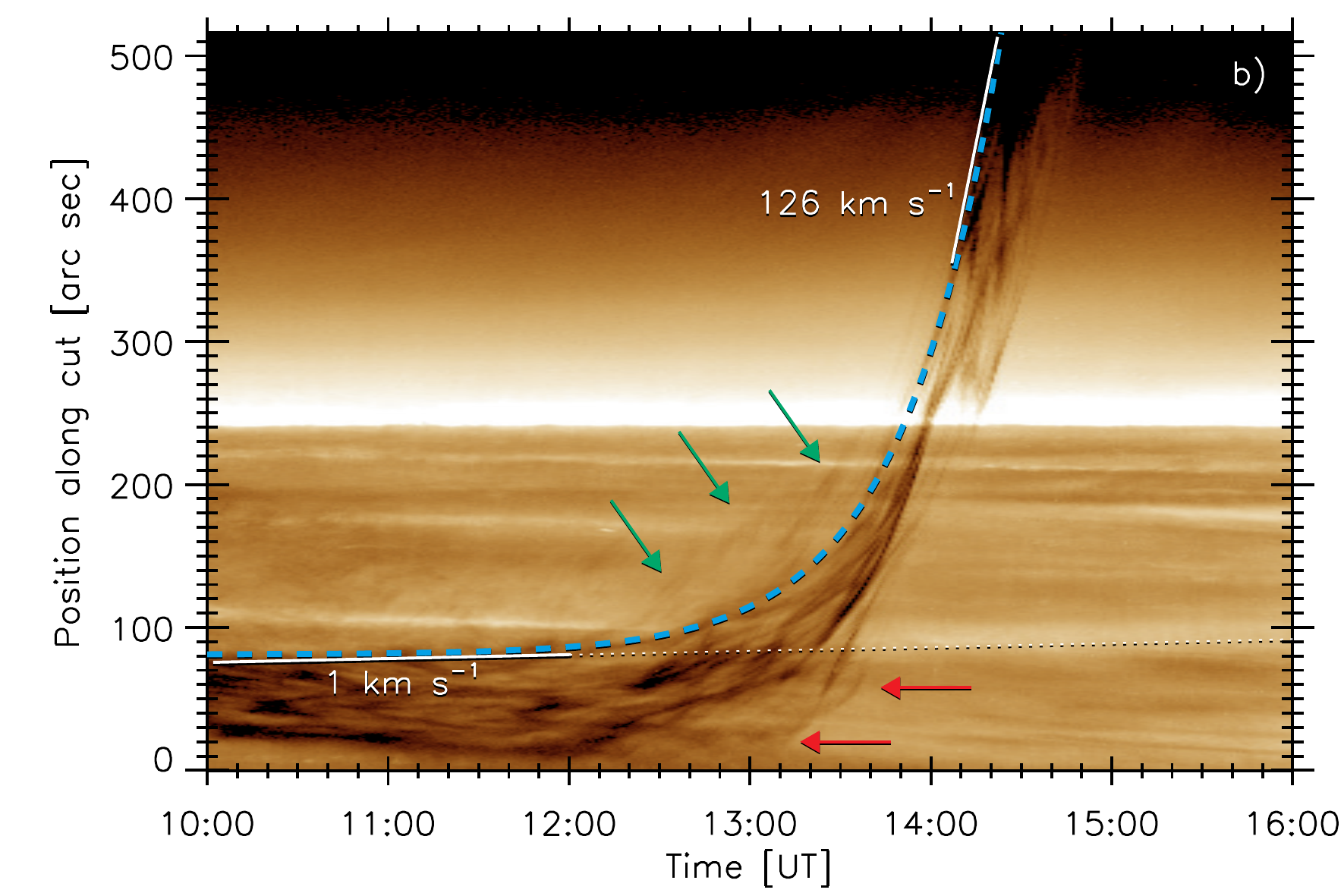}
  \caption{\textit{Left:} The eruption viewed in the 193\,\AA~filter channel of AIA. The white cut {in the \textit{left} panel} denotes the direction of the eruption and was used to produce the time-distance diagram shown in the \textit{right} panel. There, the blue dashed line is the exponential fit of the leading edge of the filament. White solid lines are fits of the slow-rise and fast-rise phases of the eruption with respective velocities. White dotted line is the extension of the linear fit of the slow-rise phase plotted for purposes of finding the onset time of the fast-rise phase. Green and red arrows mark faint threads pushed and dragged with the erupting filament, respectively. \label{fig_xt_rising}}
\end{figure*}

AIA data were processed using the standard \texttt{aia\_prep.pro} routine, normalized to exposure times, and corrected for stray light using the deconvolution method of \citet{poduval13}. Frames showing on-disk regions were derotated to a reference time of 19:00 UT {2015 April 28th} and then interpolated to a grid of stable coordinates. 

The imaging observations from AIA are supplemented with coronagraphic observations carried out using the LASCO C2 coronagraph. The LASCO data were obtained from the Virtual Solar Observatory (VSO), corrected for flat-field, and co-registered {to common} spatial coordinates. We also use measurements of the line-of-sight component of the photospheric magnetic field ($B_{\text{LOS}}$) from the Helioseismic and Magnetic Imager \citep[HMI;][]{scherrer12} onboard SDO. 

\section{Observations} \label{sec_observations}

\subsection{Overview of the 2015 April 28th filament eruption} \label{sec_eruption}

We focus on the eruption of a long filament F1. The event is introduced in Figure \ref{fig_overview}, which shows the field-of-view (FOV) of AIA in the 304\,\AA~filter channel data before (panel (a)), during (b), and after (c) the eruption, respectively. An extensive quiet Sun region is located in the east, with three filaments F1, F2, and F3 (panel (a)). In the neighborhood of the eruption lies a coronal hole (CH) and the NOAA 12333 active region (panels (a) and (b)). 

The filament F1 had been forming during nearly two Carrington rotations since early March 2015. Its eruption began after the eruption of the westernmost filament F3, which took place at $\approx$3:00 UT on 2015 April 28th. Stability of the filament F2 did not seem to be altered between the eruptions of F1 and F3. The sympathetic nature of this eruption, if any, is however out of scope of this study and we therefore do not analyse F2 and F3 any further. 

Due to its orientation and large size, the filament F1 cannot be traced along its whole extent. Using the standard solar coordinates, its southern leg was rooted in an area corresponding to $X \approx -400\arcsec$ and $Y \approx$ 200\arcsec--400\arcsec. Its northern footpoints were not visible because they were either obscured by the filament itself or located behind the solar limb. F1 {lay} above the polarity inversion line (PIL) which we roughly {indicate} using the yellow line in Figure \ref{fig_overview}(d).

Panels (e)--(g) of Figure \ref{fig_overview} show the eruption as viewed by the LASCO C2 coronagraph. In panel (e), the apex of the filament, visible as a bright blob above the occulter is overlaid by an arcade of long coronal loops. This envelope is being pushed outwards by the erupting filament, which is depicted in panel (f). There, its two legs can be distinguished, one rooted toward the north and the other toward the south. The legs are better visible in panel (g), in which most of the erupting filament has already left the FOV. Panels (f)--(g) also reveal the presence of a crossing of the filament's legs, visible between the filament and the occulter, creating a so-called writhed elbow {\citep{zhang12, chengzhang13}}. The eruption of F1 was not accompanied by a clearly identifiable flare; the \textit{GOES} hard X-ray flux was, during the whole studied period below the C-class threshold and did not show a discernible peak (panel (h)), see also Section \ref{sec_arrf}). 

\subsection{Phases of the eruption}

The eruption of F1 was a very-long lasting event, since the eruption-related phenomena were observed for more than 24 hours (see Section \ref{sec_longterm}). The filament F1 started to erupt after $\approx$12:00 UT, although motions of the threads composing its southern leg could be discerned even at earlier times. To specify the onset time of the eruption, we constructed a time-distance diagram along the direction of the eruption (white arrow in Figure \ref{fig_xt_rising}(a)). F1 was first rising slowly for about two hours up until \mbox{$\approx$12:00 UT} (Figure \ref{fig_xt_rising}(b)), when it exponentially accelerated. To distinguish between the slow- and the fast-rise phases of the eruption, we fitted the motion of the leading edge of the filament using the fitting function of \citet{cheng13}:
\begin{equation}
h(t)=c_0e^{(t-t_0)/\tau)} + c_1(t-t_0) + c2
\end{equation}
There, $c_0, c_1, c_2, t_0$ and $\tau$ are free parameters which initial estimates were taken from \citet[][Table 3 therein]{mccauley15}. The parameter $c_1$, handling the increment of the linear term, was found to be very small, of the order of 10$^{-4}$, with the uncertainty of the same order of magnitude. We have therefore discarded the linear term and fitted the filament using the exponential and constant terms only. The resulting fit is shown using the blue dashed line in Figure \ref{fig_xt_rising}(b). Separate linear fits (solid white lines) were then used to obtain the velocities of rising during the slow- and the fast-rise phases of the eruption. These were found to be \mbox{$\approx$1 km\,s$^{-1}$} and \mbox{126 km\,s$^{-1}$}, respectively.

Since we weren't capable of calculating the onset time using the {full analytic formula as in \citet{mccauley15}}, for its estimation we used the absolute differences between the linear fit of the slow-rise phase (extended using the white dotted line in Figure \ref{fig_xt_rising}(b) and the exponential fitting function. We found that these functions exceed the threshold corresponding to the spatial resolution of AIA after $\approx$12:10 UT. Similar onset time can also be obtained via expanding the fitting function into the Taylor polynomials.    

Note that this fit of the leading edge of the filament is not representative of the motion of the individual threads composing it. As can be seen in Figure \ref{fig_xt_rising}(b), at the beginning of the eruption, some threads are seen to slightly descend between $\approx$10:00 and 11:30 UT, most likely due to twisting motions of filament material during the eruption. Apart from this, faint threads can be distinguished both in front (green arrows) and behind the filament (red arrows). The transition between the slow- and the fast-rise phases for these threads occurs at various instants between $\approx$12:30 and 13:40 UT. 

\begin{figure*}[t]
  \centering    
    \includegraphics[width=5.60cm, clip,   viewport=  00 40 265 320]{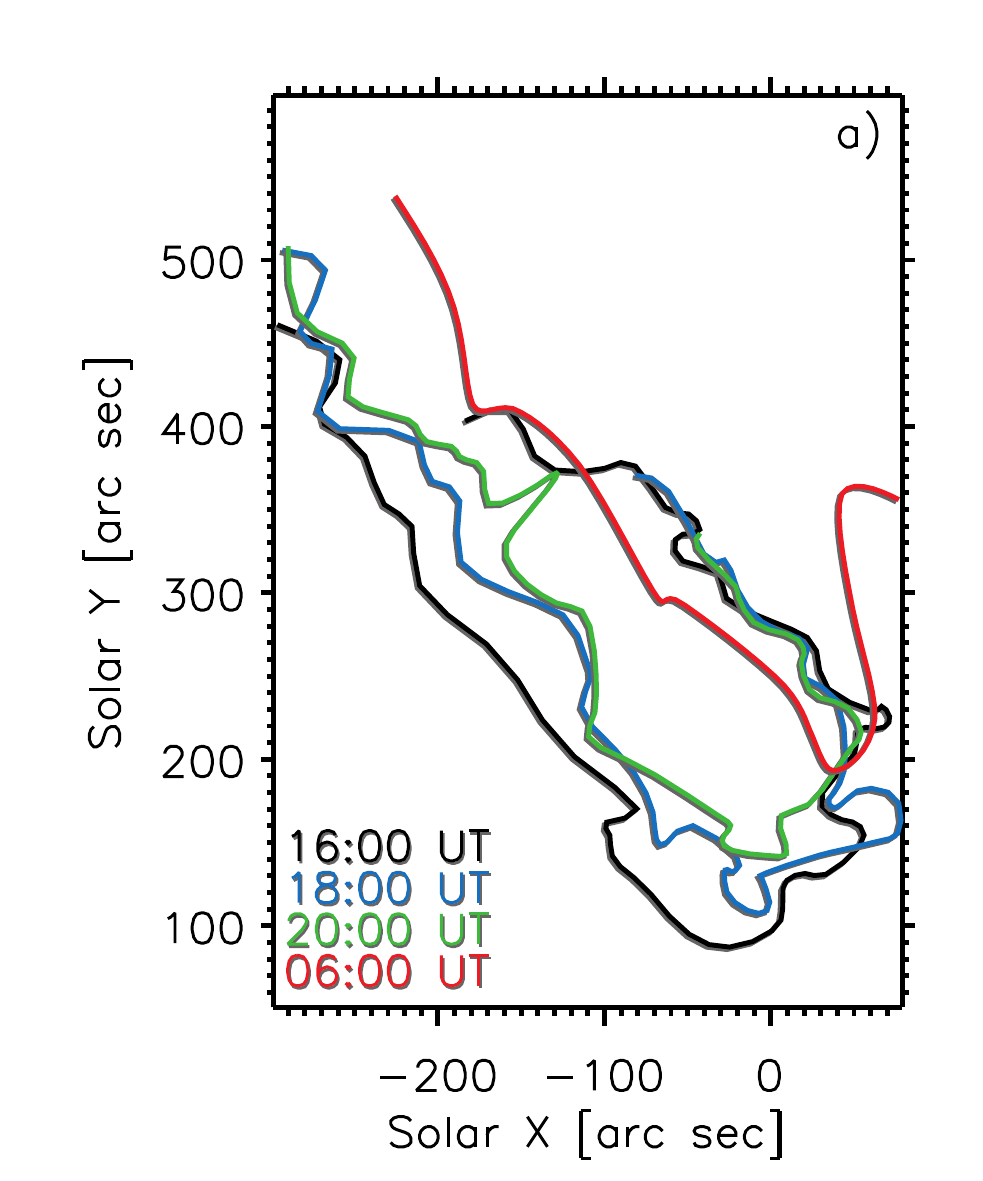}
    \includegraphics[width=4.02cm, clip,   viewport=  75 40 265 320]{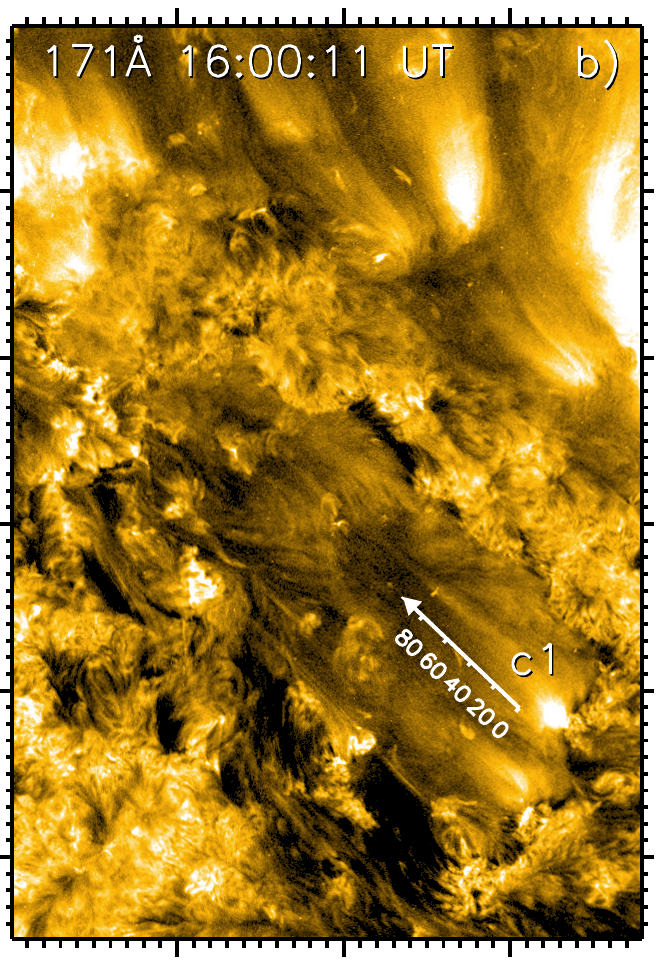}
    \includegraphics[width=4.02cm, clip,   viewport=  75 40 265 320]{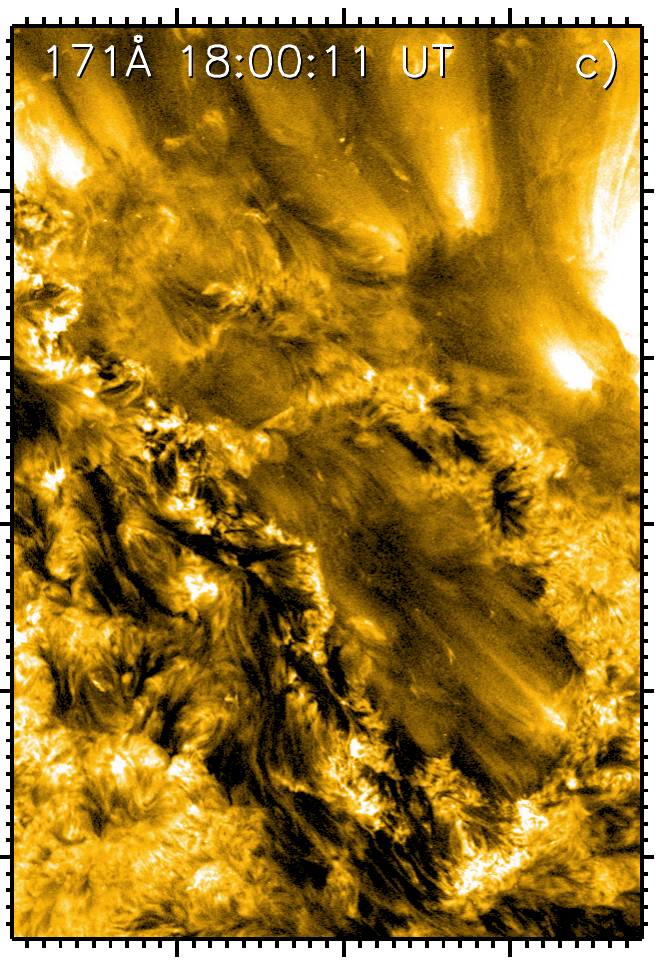}
    \includegraphics[width=4.02cm, clip,   viewport=  75 40 265 320]{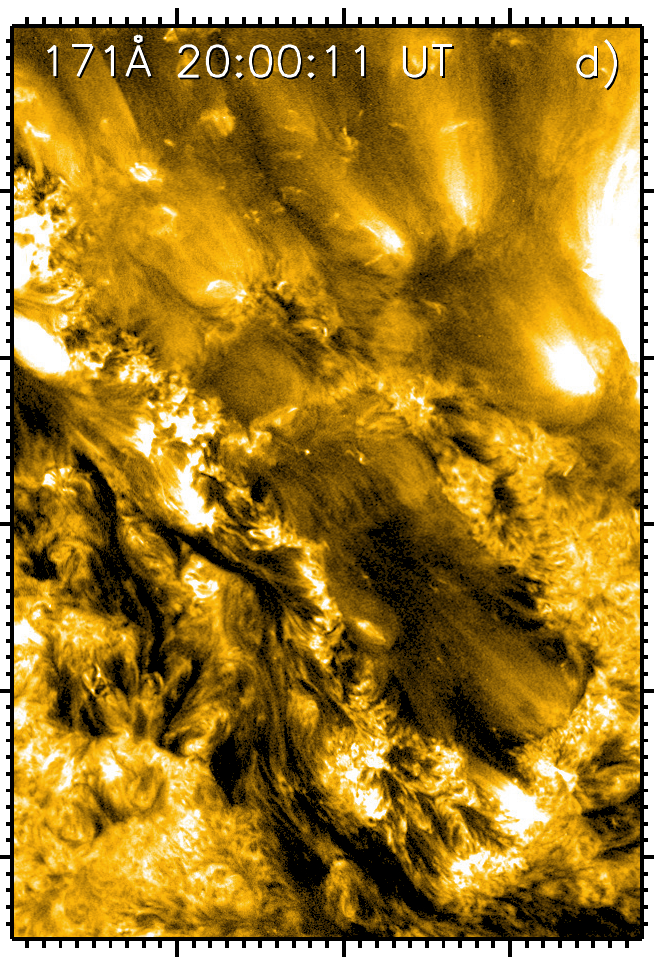}
    \\
    \includegraphics[width=5.60cm, clip,   viewport=  00 00 265 320]{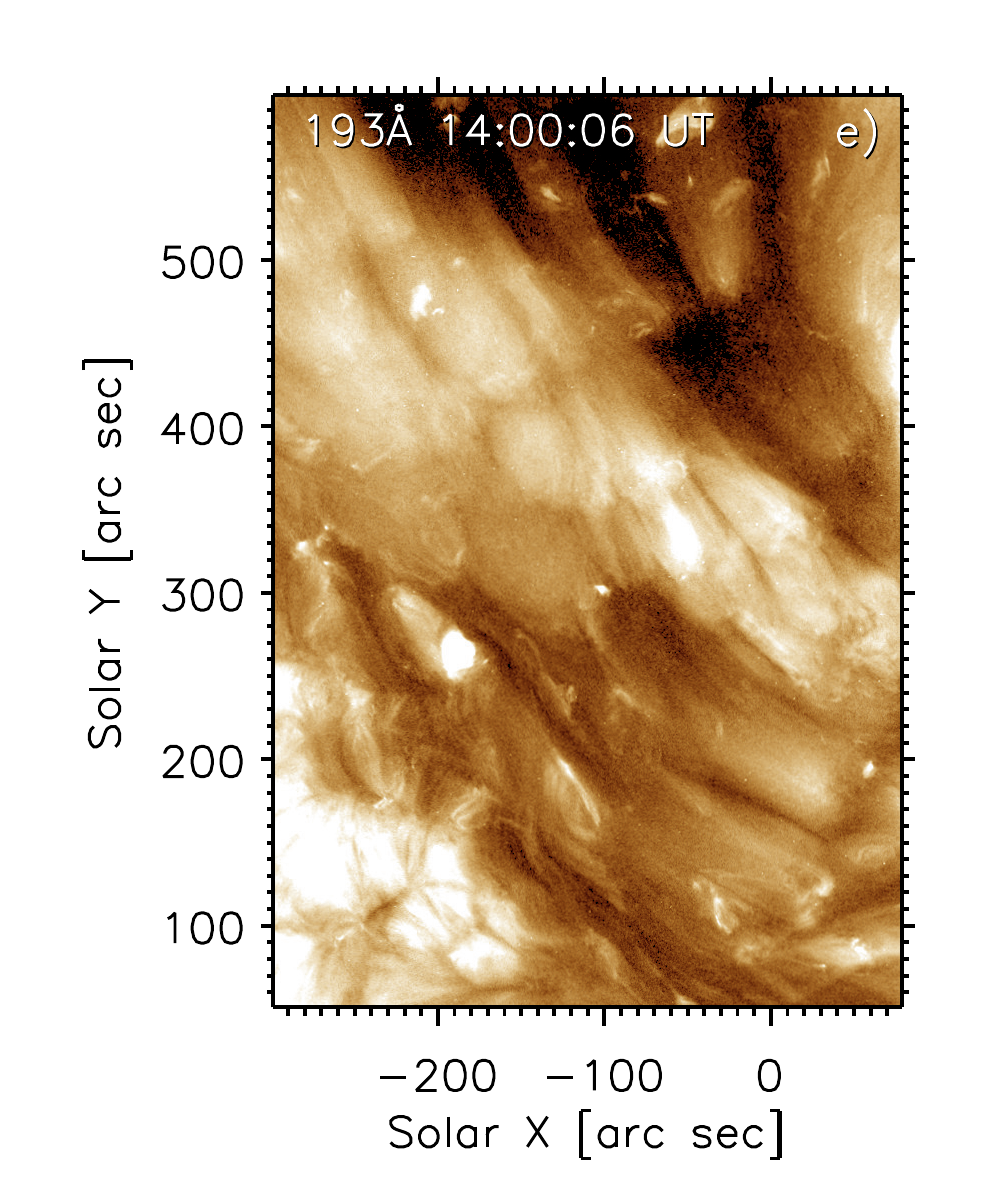}
    \includegraphics[width=4.02cm, clip,   viewport=  75 00 265 320]{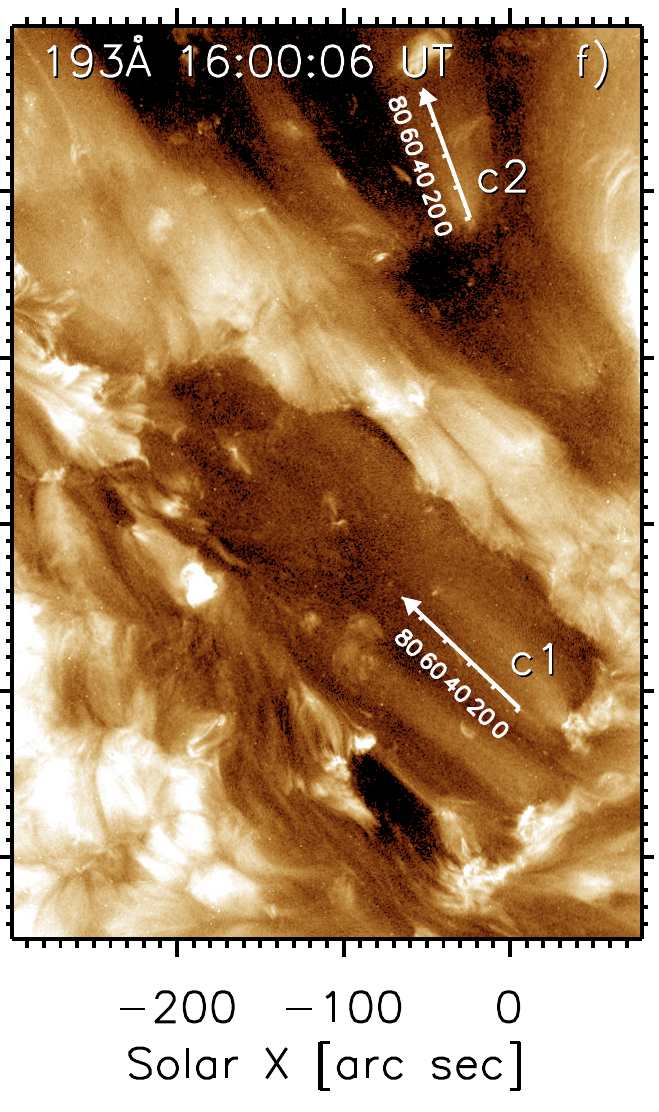}
    \includegraphics[width=4.02cm, clip,   viewport=  75 00 265 320]{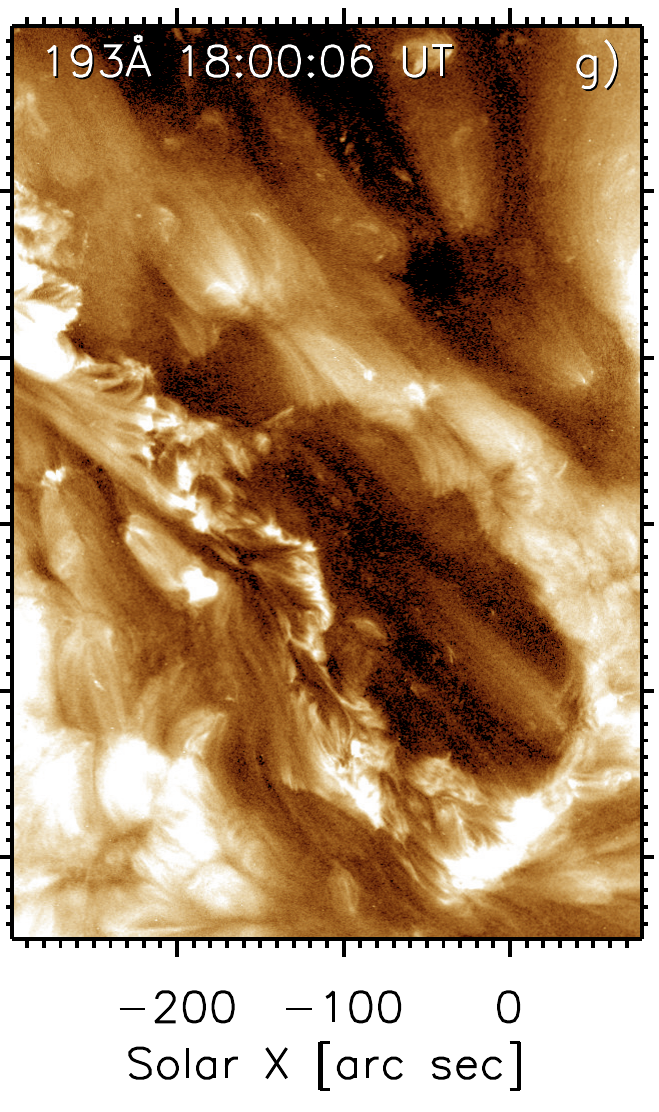}
    \includegraphics[width=4.02cm, clip,   viewport=  75 00 265 320]{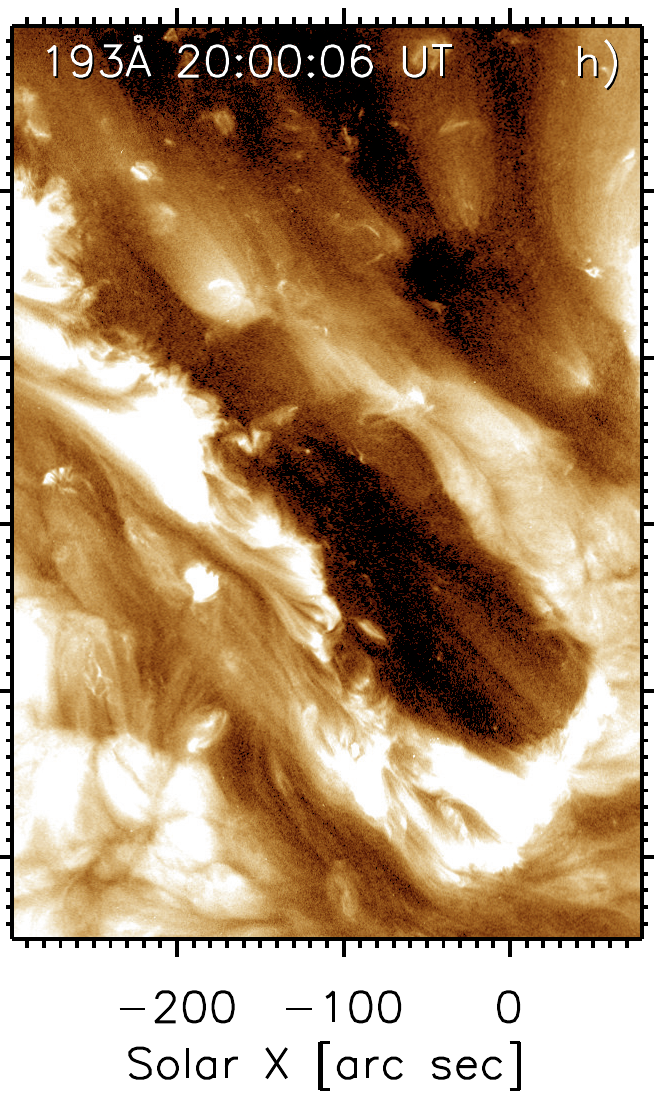}
  \caption{Overview of the positive-polarity ribbon hook. Panel (a) shows its evolution traced by a manual selection of its edges using the 304\,\AA~filter channel data. Panels (b)--(d) show the hook as viewed in the 171\,\AA~filter channel of AIA, while panels (e)--(h) show the hook in the 193\,\AA~channel. Cuts $c1$ and $c2$ in panels (b) and (f) are directd along the funnels analysed for a presence of flows in Sections \ref{sec_flows_dr} and \ref{sec_flows_ch}, respectively.\\ {Animated version of this figure is available in the online journal. The animation covers the period of 14:00 UT -- 21:00 UT, while its real-time duration is 20 seconds.} \label{fig_hook}}
\end{figure*}

\subsection{Flare ribbons} \label{sec_hookform}

During the eruption, two flare ribbons formed and at least one of them developed a $J$-shaped (hooked) extension (Figure \ref{fig_overview}(c)). In order to investigate {which magnetic polarities were occupied by the ribbons}, we manually traced the ribbons using the 304\,\AA~filter channel data and re-plotted them over HMI $B_{\text{LOS}}$ data saturated to $\pm$ 30 G (red dashed lines in Figure \ref{fig_overview}(d)). Since the {eastern} ribbon is spatially coincident with regions of the negative magnetic polarity, we will refer to it as the negative-polarity ribbon. Analogically, {we will refer to the conjugate hooked ribbon rooted in the positive magnetic polarity as} the positive-polarity ribbon and to its hook as `the hook'. 

The evolution of flare ribbons and in particular that of the hook is here first traced manually using the 304\,\AA~filter channel data. In this channel, the ribbons are well-visible and unobstructed by the coronal emission. Positions of the positive-polarity ribbon and the hook at different times are shown in Figure \ref{fig_hook}(a). The hook started to form roughly after 14:00 UT, during the fast-rise phase of the eruption (Section \ref{sec_eruption}). We were first able to distinguish its curved part (elbow) located in the south. Both its ends then elongated toward northeast {(along the PIL)}, one extending to what became the tip of the hook, while the other towards the straight part of the positive-polarity {ribbon. At this point, the} entire hooked ribbon became visible as a single structure (black line). The hook then started to contract {and, together with the rest of the ribbon to which it was attached, drift toward the north (blue, green, and red lines)}. 

The hook can also be observed in the coronal filter channels. The 171\,\AA~and 193\,\AA~filter channel data are shown in panels (c)--(h) at a 2-hour cadence. Animated versions of these observations are available online. In the 171\,\AA~filter (panels (b)--(d)), quiet Sun network surrounding the hook can be seen. Panels (e)--(h) showing the 193\,\AA~filter reveal the quiet Sun region with numerous funnel-shaped loop footpoints. The relatively-darker region seen in the southern part of panel (e) is the hook in early stages of its formation. Later, after the hook elongated (panel (f)), we see that some of the funnel-shaped loop footpoints are located at the inner edge of the hook, for example at $\approx$[50$\arcsec$, 180$\arcsec$]. For simplicity, these will be further referred to as the funnels. Apart from the funnels, low-lying bundles of quiet-Sun loops are seen to surround the hook. Later, as the hook drifted toward {north}, some of these structures disappeared (panels (g)--(h)), presumably by the magnetic reconnection at the ribbon hook. This phenomenon is further analysed in Section \ref{sec_arcades}. 

\begin{figure*}[h]
  \centering    
    \includegraphics[width=22.00cm, angle=90, clip,   viewport=  0 45 907 200]{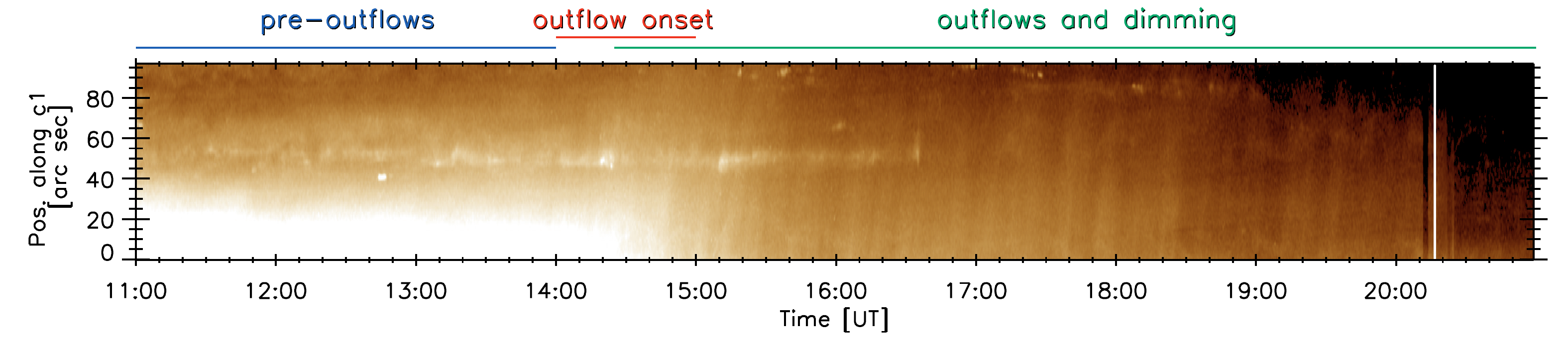}
    \includegraphics[width=22.00cm, angle=90, clip,   viewport=  0 45 907 160]{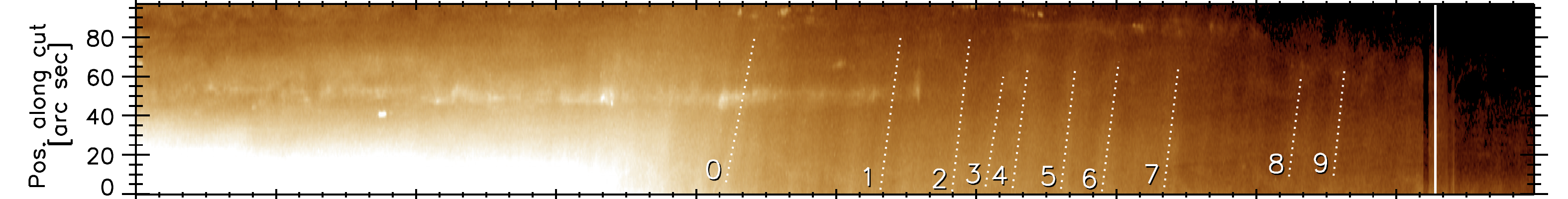}
    \includegraphics[width=22.00cm, angle=90, clip,   viewport=  0 45 907 160]{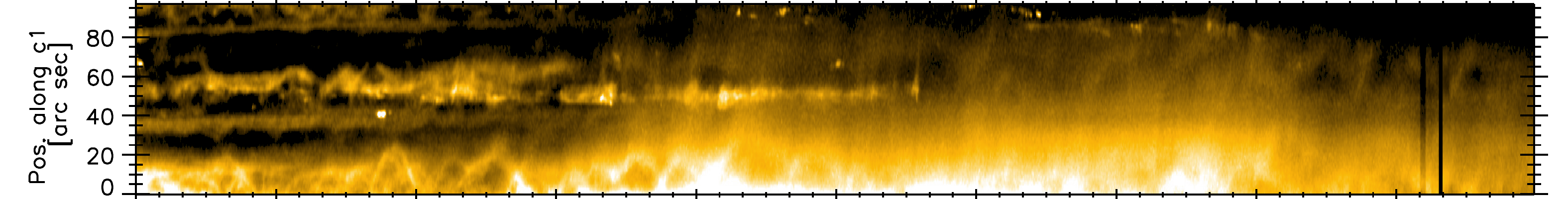}
    \includegraphics[width=22.00cm, angle=90, clip,   viewport=  0 05 907 160]{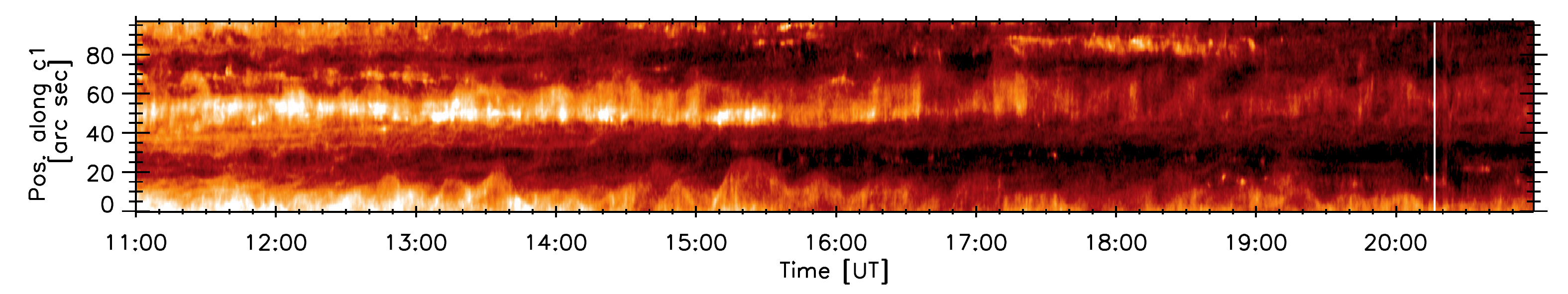}
   \caption{Time-distance diagrams constructed along the funnel in the hook ($c1$ in Figure \ref{fig_hook}). Colored lines above the figure mark three phases of the outflows. Numbered dotted lines in the second 193\,\AA~time-distance diagram are linear fits of stripes used to measure the velocities of the outflows listed in Table \ref{tab_vel_dr}. \label{fig_outf_hook}}
\end{figure*}

\begin{figure*}[t]
  \centering    
    \includegraphics[width=18cm, clip,   viewport= 0 0 623 220]{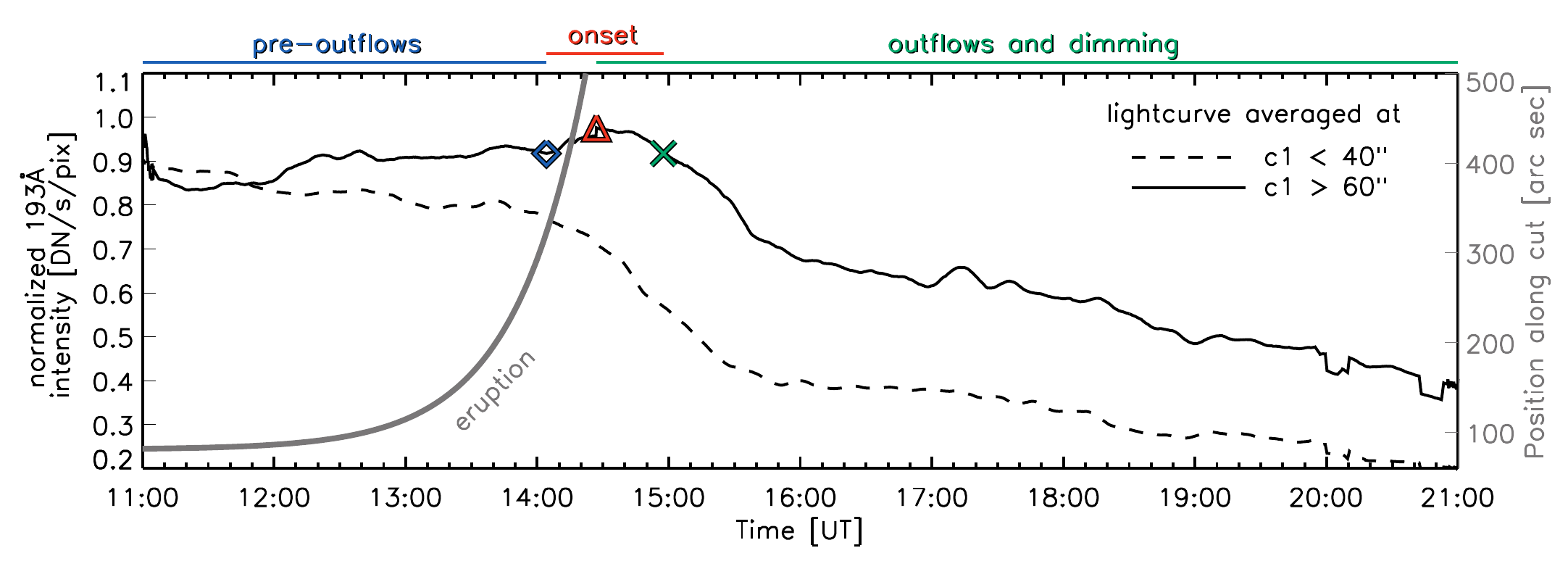}
  \caption{Lightcurves produced using the data shown in the 193\,\AA~time-distance diagram shown in Figure \ref{fig_outf_hook}. Different linestyles were used to distinguish between the positions of the cut $c1$ used for data averaging. In addition, the lightcurves were smoothed over 10-minute intervals. The blue diamond, red triangle, and green cross symbols mark the threshold we used for defining the three phases of the outflows along the analysed funnel. In analogy with Figure \ref{fig_outf_hook}, these are marked using the colored lines above the figure. The grey thick curve is the fit of the erupting filament from Figure \ref{fig_xt_rising}. \label{fig_outf_hook_suppl} }
\end{figure*}

\subsection{Formation of core coronal dimming} \label{sec_dimmform}

After the eruption, a coronal dimming formed in the area encircled by the hook (Figure \ref{fig_hook}(f)--(h)). The intensity drop was caused by the disappearance of the individual structures rooted in this region, mainly of the funnels which were progressively narrowing (panels (b)--(d) and (f)--(h)) until they finally faded away after 21:00 UT. There were also multiple funnels rooted in the northern part of the dimming region, which faded within 2 hours after its formation (panels (e)--(f)). Besides, a few small-scale transition region (TR) structures were observed both in the 171\,\AA~and 193\,\AA~filters. Most of them disappeared before 20:00 UT, as shown in Figure \ref{fig_hook}.

\section{Plasma outflows} \label{sec_outflows}

The animation accompanying Figure \ref{fig_hook} reveals motions of plasma along most of the funnels located in and near the dimming region. In the following section, we analyse their typical velocities and we investigate whether they are associated with other phenomena related to the eruption.

\subsection{Outflows from the dimming region} \label{sec_flows_dr}

Funnels located at the inner edge of the hook are projected to lean over the dimming region, which naturally enhances their contrast. We investigate the funnel marked using the cut $c1$ in Figure \ref{fig_hook}(f), because it is both the largest and the brightest one and is visible for the longest period. Time-distance diagrams constructed along this funnel are shown in Figure \ref{fig_outf_hook}. They contain full-cadence data imaged in the 193\,\AA, 171\,\AA, and 304\,\AA~filter channels of AIA. To increase the signal-to-noise ratio, they were smoothed with a \mbox{3\,$\times$\,3} pixel boxcar, which roughly corresponds to the spatial resolution of the instrument.

The time-distance diagram showing the 193\,\AA~data is first dominated by a bright region at the position $c1 \lessapprox25\arcsec$. This emission originates in the lower portions of the funnel and fades after 14:20 UT. Roughly at the same time, the funnel seems to slightly brighten along the whole extent of $c1$. This phase remains present till $\approx$15:00 UT, when vertical stripes, repeated in a rib-like pattern, start to be visible along most of $c1$. The inclination of these stripes indicates a presence of a motion directed along $c1$. Given the orientation of $c1$, this implies that the material is moving from lower to higher altitudes. This phenomenon lasts for more than five hours till $\approx$20:20 UT. During this time, the funnel gradually fades until it disappears (see Figure \ref{fig_hook}). In the remainder of this section, we will assume this motion to be a signature of outflows, while a discussion on its nature is left for Section \ref{sec_interpretation}.

\begin{figure*}[t]
  \centering    
    \includegraphics[width=11.00cm, clip,   viewport= 0 40 453 170]{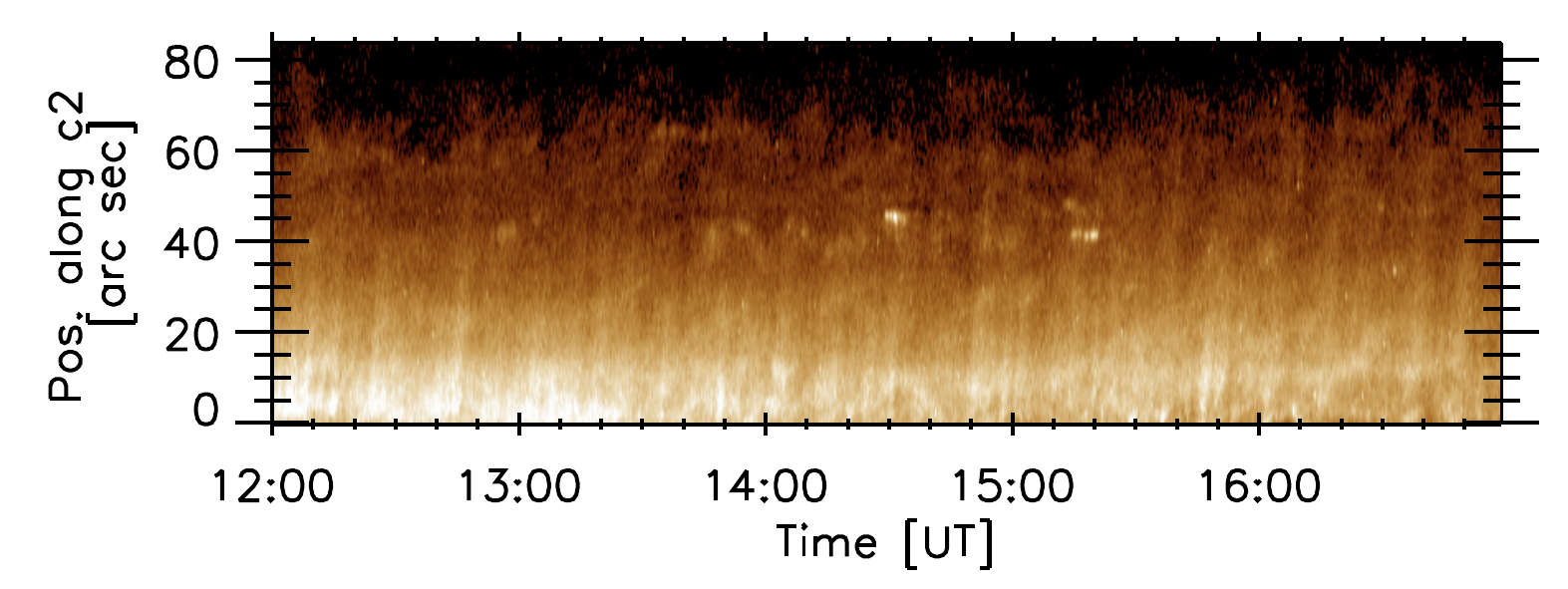}
    \\
    \includegraphics[width=11.00cm, clip,   viewport= 0 0 453 170]{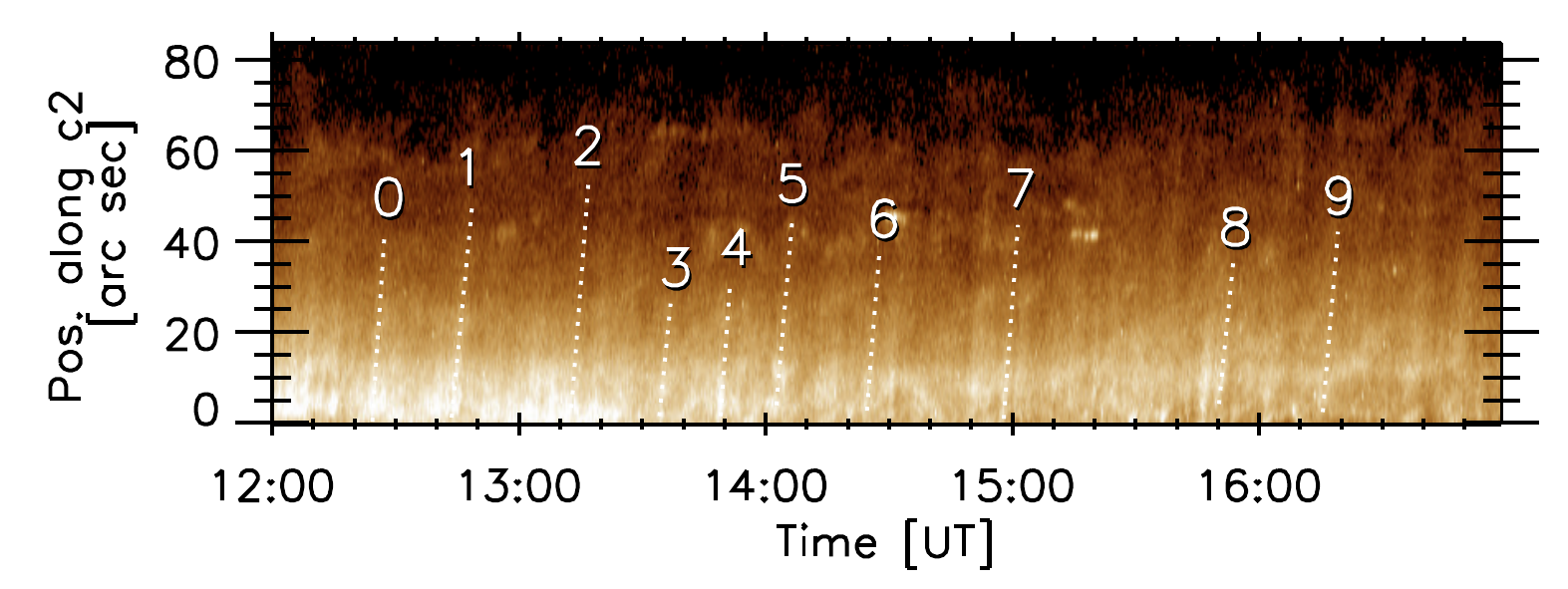}
  \caption{Time-distance diagrams evidencing the presence of the outflows along the funnel in the coronal hole. They were produced using the 193\,\AA~filter channel data along the cut $c2$. Numbered dotted lines are linear fits of stripes used for derivation of the velocities of the outflows listed in Table \ref{tab_vel_ch}. \label{fig_outf_ch}}
\end{figure*}

The period shown in Figure \ref{fig_outf_hook} can, {to} a first approximation, be divided into three different episodes, named the `pre-outflows', `onset', and `outflows and dimming' phases. They are named according to the variations of the brightness of the funnel and the presence of the outflows, and were highlighted using colored lines plotted above Figure \ref{fig_outf_hook}. 

In order to find the instants limiting the {commencement and termination} of each phase, we produced lightcurves using the 193\,\AA~channel data averaged along $c1$. Normalized lightcurves, smoothed over 10-minute intervals, are shown in Figure \ref{fig_outf_hook_suppl}. The lightcurves were produced along the upper part of the cut, $c1 > 60\arcsec$, and along its lower part, $c1 < 40\arcsec$. This choice excludes the bright underlying transition region that is projected at positions \mbox{$c1 \approx40\arcsec-60\arcsec$} along the cut. In the same figure we also plot the function fitting the rise of the filament (grey line), derived in Figure \ref{fig_xt_rising}(a). 

The dashed line, showing the intensities averaged at low altitudes \mbox({$c1 < 40\arcsec$}, dashed line), is decreasing during most of the analysed period. This decrease was typically slow except for the period between $\approx$13:30 UT and $\approx$15:30 UT, when the averaged intensity dropped by about 40\%. This darkening corresponds to the formation of the hook and the dimming region within. On the contrary, the intensity averaged near the tip of $c1$ (\mbox{$c1 > 60\arcsec$}, solid line) first slightly increased and after $\approx$12:30 UT reached a plateau. After $\approx$14:00 UT it rose again, peaked at $\approx$14:30 UT and started to drop. The averaged intensity reached the initial pre-rise value at $\approx$15:00 UT. These three instants are {highlighted in Figure \ref{fig_outf_hook_suppl}} using the blue diamond, red triangle, and green cross symbols. Note that after the peak of the solid curve, both lightcurves were decreasing in a similar manner. This implies that the funnel and the outflows along it were dimming evenly along their whole extent.  

It is difficult to find the exact time of the onset of the outflows using Figure \ref{fig_outf_hook} only, as the transition is gradual and the event is very long-lasting. We will therefore refer to the period of the increased brightness, between {the blue} diamond and the green cross, as the transition phase between the `pre-outflow phase' to the phase characterized by the presence of the outflows (`outflows and dimming'). The transition phase is co-temporal with the fast-rise phase of the eruption (Section \ref{sec_eruption}) and the formation of the dimming region (Section \ref{sec_dimmform}). For simplicity, we will refer to this period as the outflow `onset' phase. 

Apart from their gradual fading, the pattern generated by the outflows in Figure \ref{fig_outf_hook} as well as their inclination do not change during the third phase. To measure the velocities of the outflows, we fitted multiple stripes observed approximately between $\approx$15:00 UT and 19:40 UT with linear fits {whose} endpoints were selected manually. The resulting velocities, listed in Table \ref{tab_vel_dr}, range between $\approx70$ and 140 km\,s$^{-1}$, while the mean velocity was 112 km\,s$^{-1}$. Their uncertainties were estimated using Equation (1) of \citet{dudik17}. There, the spatial resolution of AIA was adopted as the uncertainty in the spatial position $\sigma_{\text{s}}$. The uncertainty in time ($\sigma_{\text{t}}$) was, by measuring the FWHM of the fitted stripes, estimated as 54\,s. 

The outflows can also be seen in the time-distance diagram produced using the 171\,\AA~filter channel data. During the pre-eruption phase, the TR structures under $c1$ are seen to emit at positions $c1 \approx40\arcsec$, 55$\arcsec$, and $>80\arcsec$. Apart from that, an irregular $V$-like pattern at $c1<25\arcsec$ can be seen. It is caused by a swaying motion of the TR structures near the footpoints of the funnels and lasts till $\approx$19:00 UT. The outflows can easily be distinguished after 14:00 UT. The outflows share the same pattern and inclination as those seen in the 193\,\AA~diagram, but after $\approx$19:00 UT, their intensity decreases more slowly. 

The outflows cannot be found in the the time-distance diagram produced using the 304\,\AA~channel data. It only reveals the $V$-like pattern in the transition region, similar to the one observed in the 171\,\AA~data. Besides, vertically-stretched features at $c1 \approx40\arcsec-60\arcsec$ and $c1\gtrapprox80\arcsec$ can be seen. Most of them fade away in the outflow phase. 

\begin{deluxetable}{cccc}[t]
\tablecaption{Outflow velocities and velocity uncertainties measured in the dimming region, resulting from different fits. \label{tab_vel_dr}}
\tablecolumns{4}
\tablenum{1}
\tablewidth{0pt}
\tablehead{
\colhead{\#} & \colhead{$v\pm dv$ [km s$^{-1}$]} & \colhead{\#} & \colhead{$v\pm dv$ [km s$^{-1}$]}}
\startdata
0 & 	73$\pm$8 		& 5	& 	122$\pm$26		\\ 
1 & 	109$\pm$16 		& 6	& 	111$\pm$20		\\ 
2 & 	124$\pm$21		& 7	& 	122$\pm$26		\\ 
3 & 	89$\pm$15 		& 8	& 	119$\pm$29  	\\ 
4 & 	114$\pm$23		& 9	& 	140$\pm$37 	\\ \hline
\enddata
\end{deluxetable}

\subsection{Outflows from the coronal hole} \label{sec_flows_ch}

We now compare the outflows in the dimming region to the outflows along funnels rooted in a proper coronal hole located in the vicinity. In its southern part, three funnels can be distinguished both in the 171\,\AA~and 193\,\AA~filter channel data (Figure \ref{fig_hook}). Their characteristics are studied using the time-distance diagrams, which we constructed along the funnel highlighted using the cut $c2$ (Figure \ref{fig_hook}(f)).  

\begin{figure*}[t]
  \centering    
    \includegraphics[width=5.19cm, clip,   viewport=  22 0 300 330]{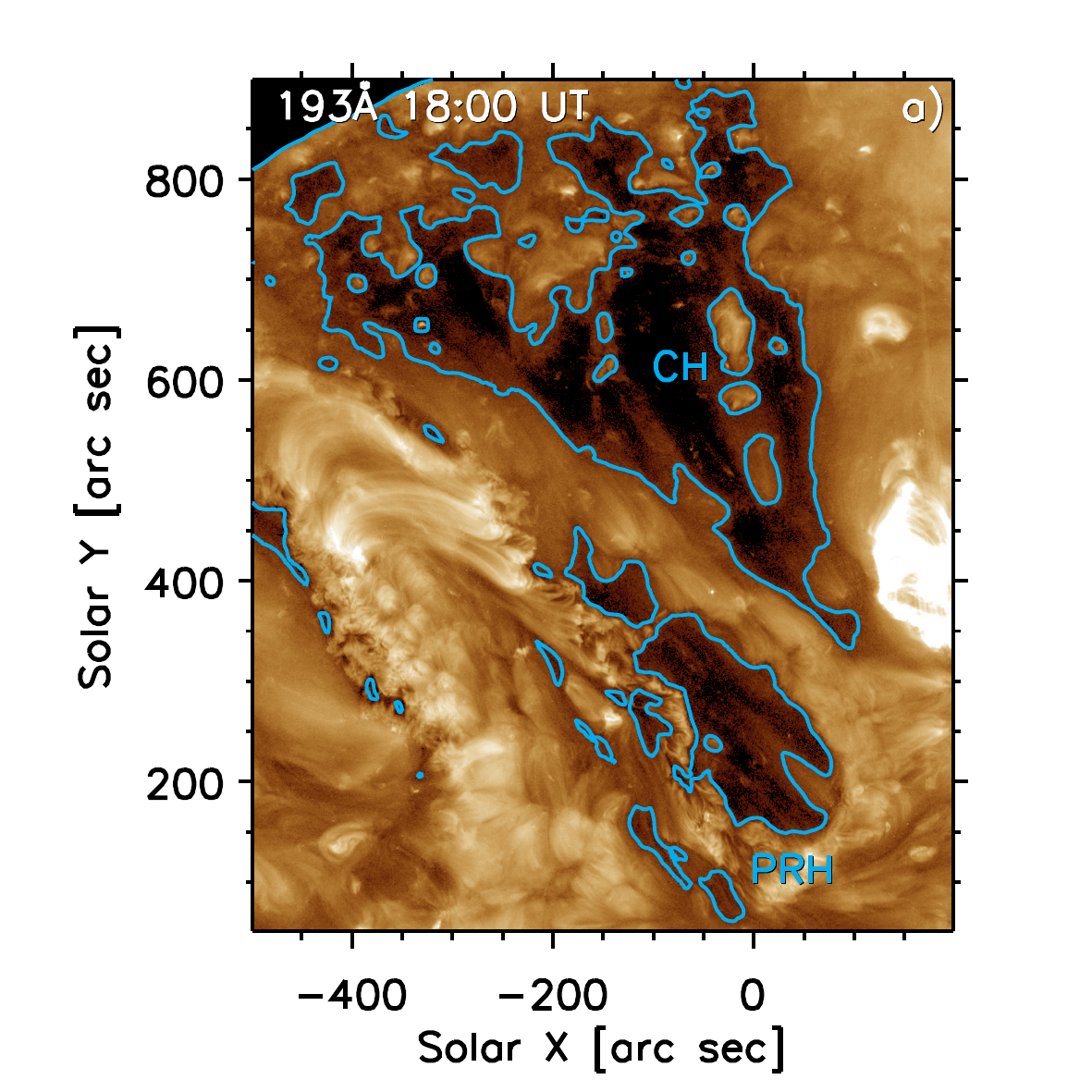}
    \includegraphics[width=4.20cm, clip,   viewport=  75 0 300 330]{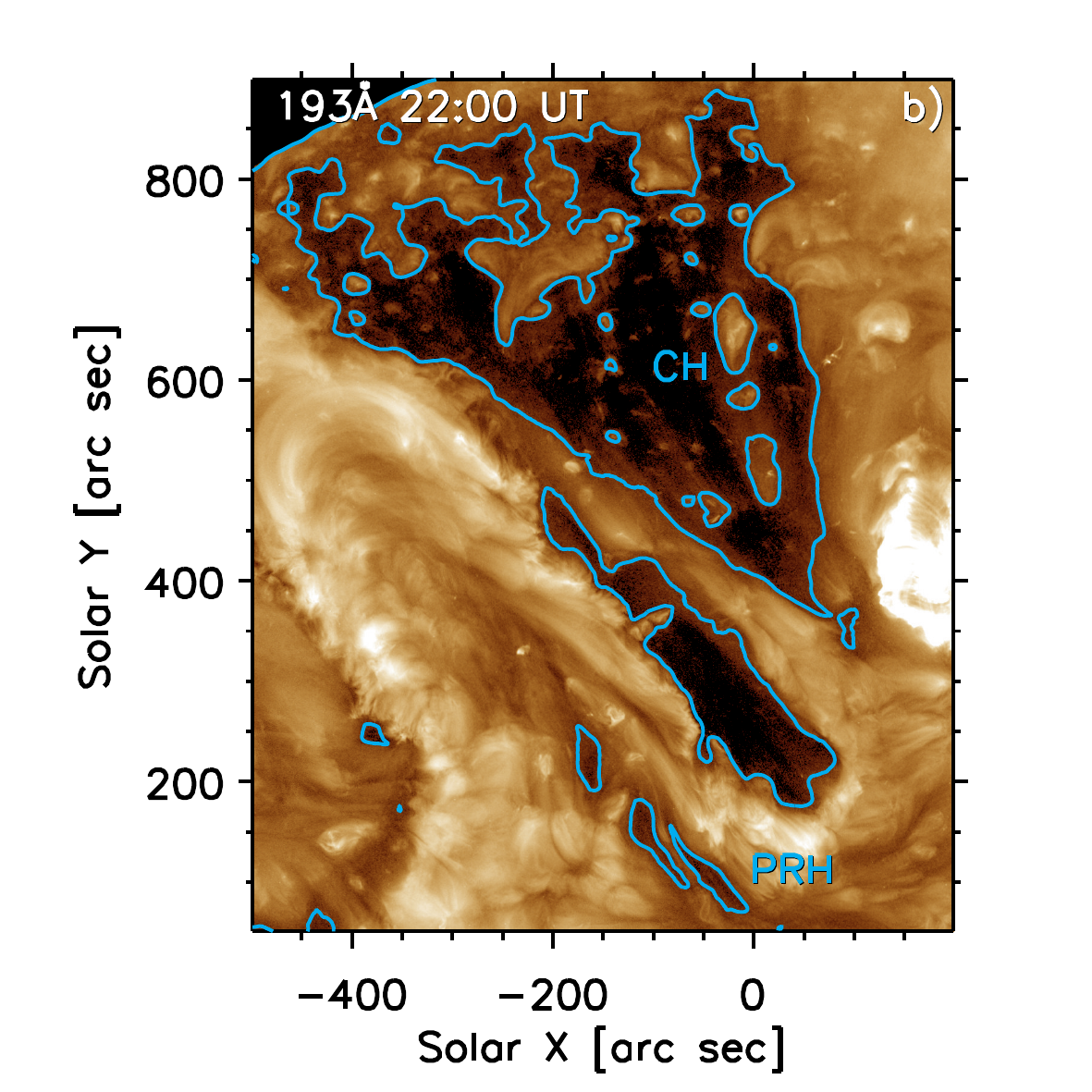}
    \includegraphics[width=4.20cm, clip,   viewport=  75 0 300 330]{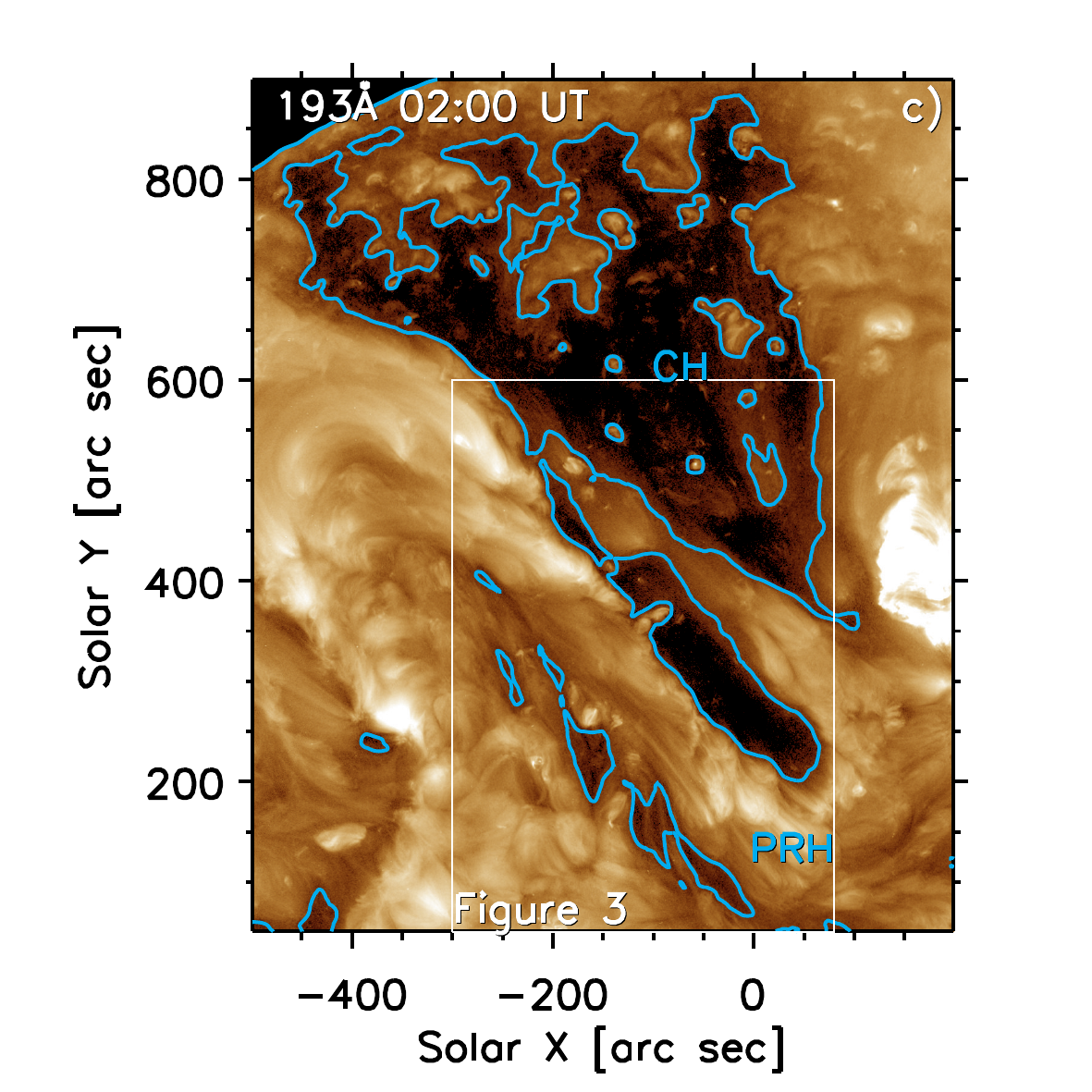}
    \includegraphics[width=4.20cm, clip,   viewport=  75 0 300 330]{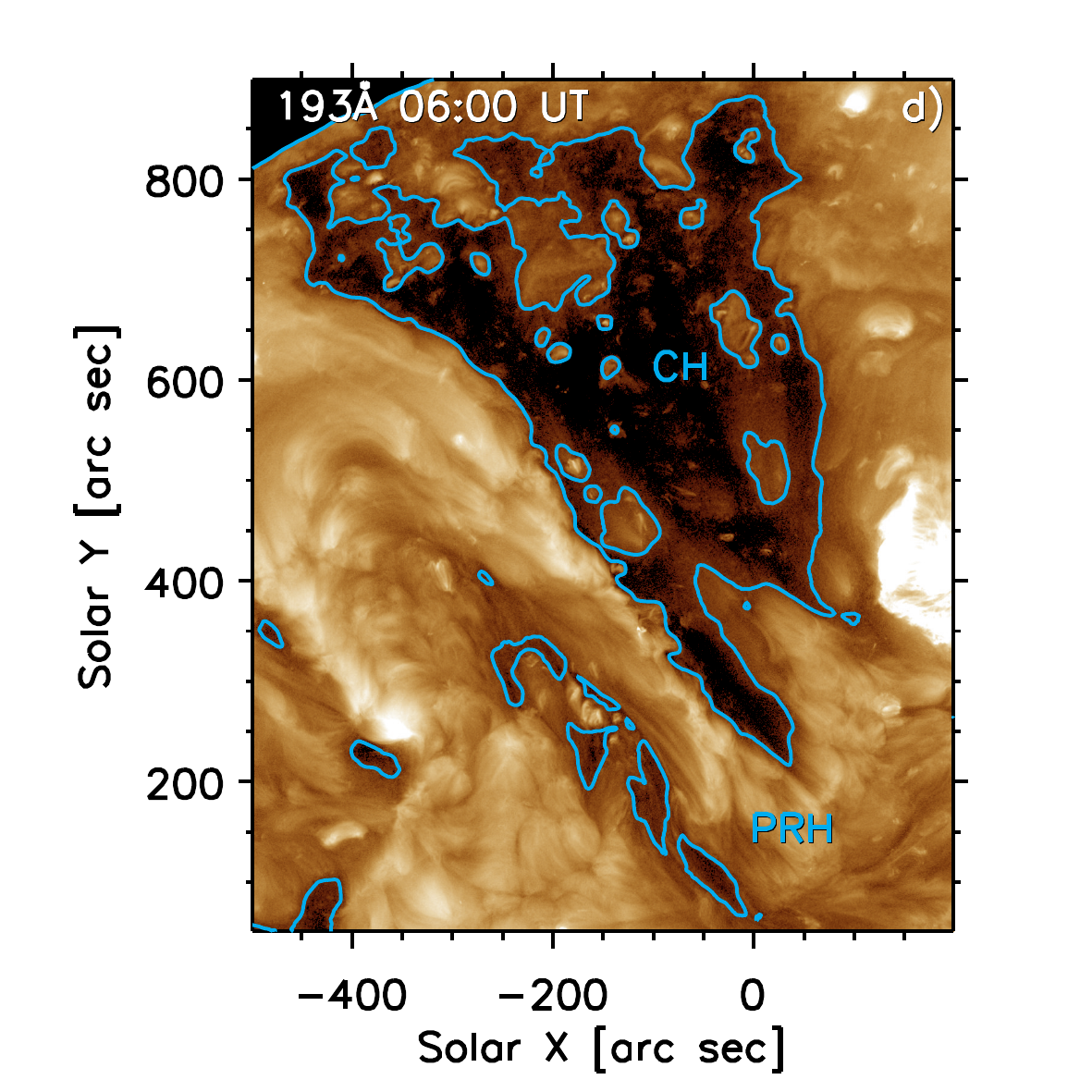}
  \caption{Merging of the hook with the coronal hole shown at a cadence of 4 hours, as viewed in the 193\,\AA{} filter channel of AIA. Blue contours correspond to 35 DN/s/pix. \label{fig_merging}}
\end{figure*} 

The time-distance diagram constructed using the 193\,\AA~filter channel data is shown in Figure \ref{fig_outf_ch}. Since the eruption had no apparent effect on the morphology of the funnels in the coronal hole, the period included therein is shortened compared to the one shown in Figure \ref{fig_outf_hook}. However, to ease the comparison between the time-distance diagrams constructed in the coronal hole and the dimming region, the one-hour intervals on the $X$-axis corresponding have the same plotted length. 

At positions $c2\lessapprox70\arcsec$, vertical, inclined stripes indicating outflows can be distinguished during the whole period. When compared to the outflows seen in the hook, they seem to be sharper and their contrast is higher, i.e., they can be recognized more easily. The stripes are steeper and narrower, the corresponding $\sigma_{\text{t}}$ is 48\,s only. Even though their mean velocity of  135 km\,s$^{-1}$ is slightly larger than in the hook, the individual velocities listed in Table \ref{tab_vel_ch} are, within the uncertainty, comparable to those reported on in the hook. We note that the outflows can also be seen in time-distance diagrams produced using the 171\,\AA~filter channel data (not shown). They are however more difficult to distinguish. 

Note that apart from the outflows, a few bright blobs are seen at $c2 = 40\arcsec - 50\arcsec$. Again, their emission corresponds to TR structures emitting from underneath $c2$. 

\subsection{Long-term evolution of the hook} \label{sec_longterm}

During and after the eruption, the positive-polarity ribbon hook with the dimming region within were slowly drifting toward the coronal hole located to the north. This convergence is shown in Figure \ref{fig_merging}, where the contours corresponding to 35 \dn were plotted to highlight the boundaries of the hook and the coronal hole. As seen in panels (b)--(c), this convergence was accompanied by fading of low-lying bundles of quiet-Sun loops rooted in the relatively-narrow region between the dimming region and the coronal hole. We note that even though their morphology is similar to those of the funnels located in the dimming region, no outflows were observed to occur along them. For simplicity, these will further be referred to as the canopies.

The following day at $\approx$06:00 UT (panel (d)) the canopies separating the hook and the coronal hole disappeared, evidencing a merging of these two structures. Positions corresponding to the hook selected by hand at this time have also been indicated in Figure \ref{fig_hook}(a) using the red line. There, opening of the hook's northern part toward the coronal hole is evident. After $\approx$16:00 UT on April 29, traces of the hook can no longer be distinguished. 

Note that this observation of merging of a {dimming region with an ordinary coronal hole} is not unique. For example, this phenomenon can be seen to follow multiple filament eruptions studied by \citet[][see Figure 2(d)--(g) therein]{gutierrez18}.

\section{Interpretation} \label{sec_interpretation}

In this section, we further discuss the characteristics of the outflows and attempt to address the physical mechanism leading to their presence during and after the filament's eruption. 

\subsection{Outflows vs. waves}

As mentioned in Section \ref{sec_intro}, imaging observations evidencing the presence of the outward-oriented plasma motions from dimming regions have not been reported so far. They are however often observed in active regions where they are referred to as the propagating intensity, or coronal, disturbances \citep[see e.g., the review of][]{demoortel09}. Even though these structures are morphologically distinct from dimmings, their edges are often blue-shifted with outflow velocities corresponding to those found in dimming regions (Section \ref{sec_intro}). For example, the pattern visible in Figure \ref{fig_outf_hook} is reminiscent of that reported by \citet[][see Figure 2b therein]{sakao07}. Their observations were interpreted as outflows contributing to the slow solar wind. The same event was revisited e.g., by \citet{nishizuka11} who discovered that the properties of the outflows change with height, being suggestive of a presence of upward-propagating waves along the outflows. The debate about the origin of these disturbances continued in numerous publications, for a thorough review we refer the reader e.g., to \citet{demoortel12}. An attempt to resolve the dispute about the origin of the propagating disturbances was carried out by e.g., \citet{demoortel15}. Even though they introduced a method based on observations of the Doppler perturbations as a probe to distinguish between the wave- and flow-like interpretation, they concluded that distinguishing between the two might not be possible at all. 

Concerning the eruption analysed in this manuscript, since the material was propagating along the funnel-shaped loops (i.e., along the field lines), the waves acting there would most likely be the Alfv\'en waves. If we, for simplicity, take $B_{\text{LOS}}$ to range between 1 -- 30\,G, which can be taken as the upper limit of $B_{\text{LOS}}$ (Figure \ref{fig_overview}(d)) and assume typical quiet-Sun densities of $2-3 \times 10^8$ cm$^{-3}$ \citep[e.g.,][]{delzanna18, lorincik20}, we obtain Alfv\'en speeds of the order of 10$^2$ -- 10$^3$ km\,s$^{-1}$. Their lower limit might thus correspond to the velocities of the upward-propagating brightenings in the dimming region (Table \ref{tab_vel_dr}). Some of their properties however turned us to favour the outflow scenario.

First, the variations of $B(z)$ and $N_{\text{e}}(z)$ with height would lead to the variations of $v_{\text{A}}$ along $c1$, which we did not observe. Second, the characteristics of MHD waves should change in time. Their amplitudes undergo damping \citep{demoortel03} caused mainly by the thermal conduction and viscosity \citep{mandal16, wenzhi16}. Changes in their periodicity have also been evidenced \citep{bryans16}. The properties of the outflows we report on however do not seem to change during the whole observation, as the inclination of the stripes visible in Figure \ref{fig_outf_hook} does not change for at least 5 hours. Third, even if we neglect the damping and changes of periodicity, $v_{\text{A}}$ would increase in time due to the drop in density during the evolution of the dimming \citep[see e.g.,][]{cheng12, vanninathan18, hou20}. Reported drops of density reaching up to 70\% would cause $v_{\text{A}}$ to progressively increase by a factor of up to $\approx$1.7, which is again contradicted by our observations.

\subsection{Dilution and the solar wind} \label{sec_dilution}

In the framework of the standard solar flare model and its 3D extensions, outflows in the dimming region, hosting the legs of the erupting flux rope (in our case the filament F1) should occur due to the expansion of the erupting structure, which leads to elongation of the legs of the CME. The relatively-long period during which the outflows are observed during this event however suggests that they are not related to a stretching-induced dilution of plasma only. Combining the estimated height of the erupting filament ($\approx$150 Mm) with the measured speed of the CME (126 km\,s$^{-1}$, Section \ref{sec_observations}) results in the dilution timescale of $\approx$20 minutes only. This estimate corresponds, within the order of magnitude, to the duration of the brief intensity decrease visible after the onset phase in Figure \ref{fig_outf_hook_suppl}. Still, it is by a factor of $\gtrsim 15$ shorter than the duration of the outflow phase in the dimming region. This means that the outflows in dimming regions persisting for such large timescales are likely to be powered by a different mechanism. 

The similarity between the characteristics of the flow-like motions observed in the dimming region and the coronal hole suggests that we are observing solar-wind like outflows. As the dimming region started to form, the funnel-shaped loop footpoints turned to true coronal funnels, known as one of the sources of the solar wind \citep[e.g.,][]{marsch99, tu05, marsch08, he10}. Possibly, the opening of the magnetic field during the CME led to a geometry favouring a free flow of plasma towards the interplanetary space in the same manner as in the coronal holes. As we have shown, the hook merged with the coronal hole, at which point the two became indistinguishable (Figure \ref{fig_merging}). Note however that this merging was not co-temporal with the observations of the outflows in the dimming region, as it continued for many hours after the outflows disappeared (Section \ref{sec_longterm}). 
   
\begin{deluxetable}{cccc}[t]
\tablecaption{Outflow velocities and velocity uncertainties measured in the coronal hole, resulting from different fits. \label{tab_vel_ch}}
\tablecolumns{4}
\tablenum{2}
\tablewidth{0pt}
\tablehead{
\colhead{\#} & \colhead{$v\pm dv$ [km s$^{-1}$]} & \colhead{\#} & \colhead{$v\pm dv$ [km s$^{-1}$]}}
\startdata
0 & 	168$\pm$64	 	& 5	& 	127$\pm$37		\\ 
1 & 	113$\pm$25 		& 6	& 	139$\pm$53		\\ 
2 & 	146$\pm$40		& 7	& 	150$\pm$49		\\ 
3 & 	109$\pm$45 		& 8	& 	109$\pm$36  	\\ 
4 & 	154$\pm$76		& 9	& 	131$\pm$40 	\\ \hline
\enddata
\end{deluxetable}

\begin{figure*}[h]
  	\includegraphics[width=5.03cm, clip,   viewport=  00 45 300 285]{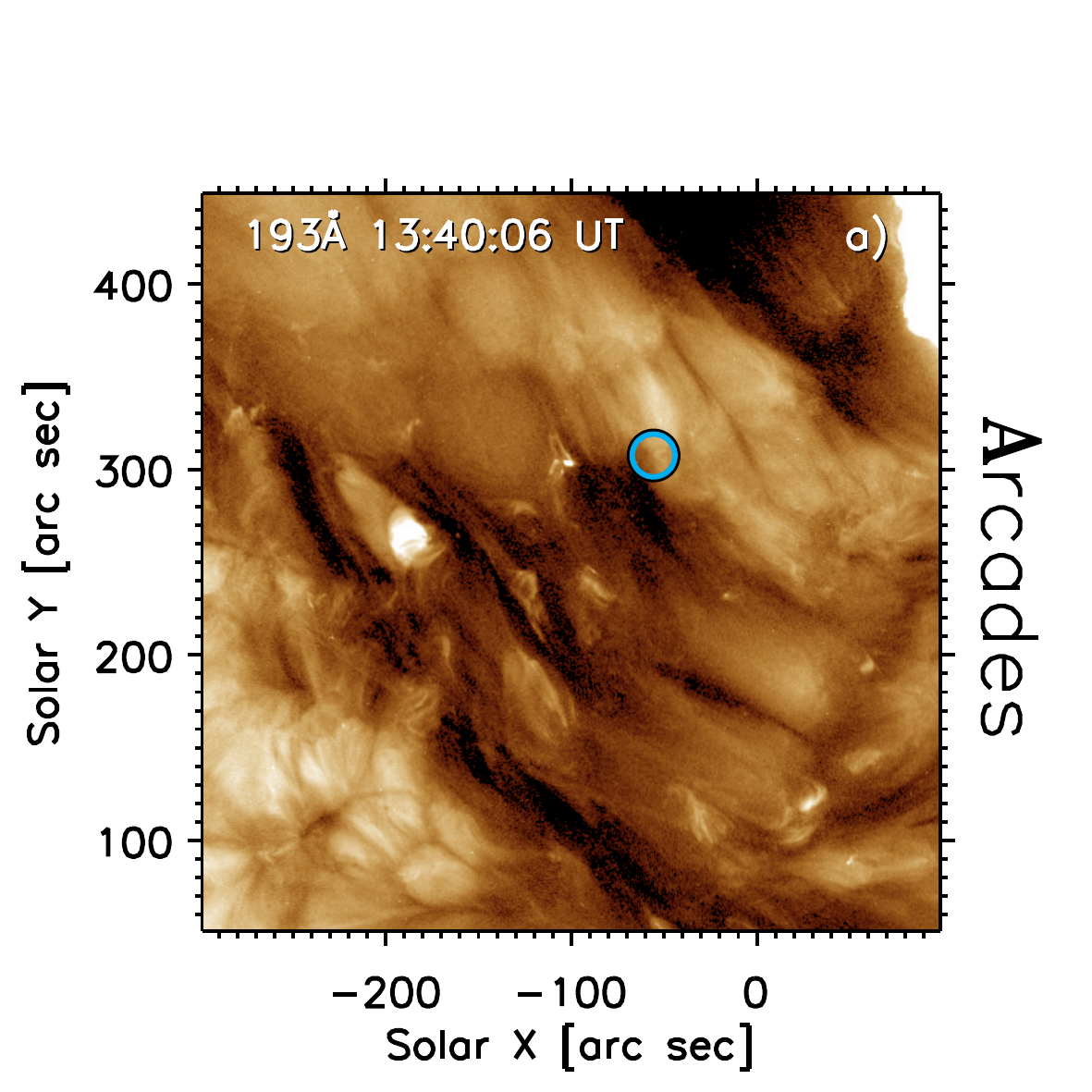}
    \includegraphics[width=4.03cm, clip,   viewport=  60 45 300 285]{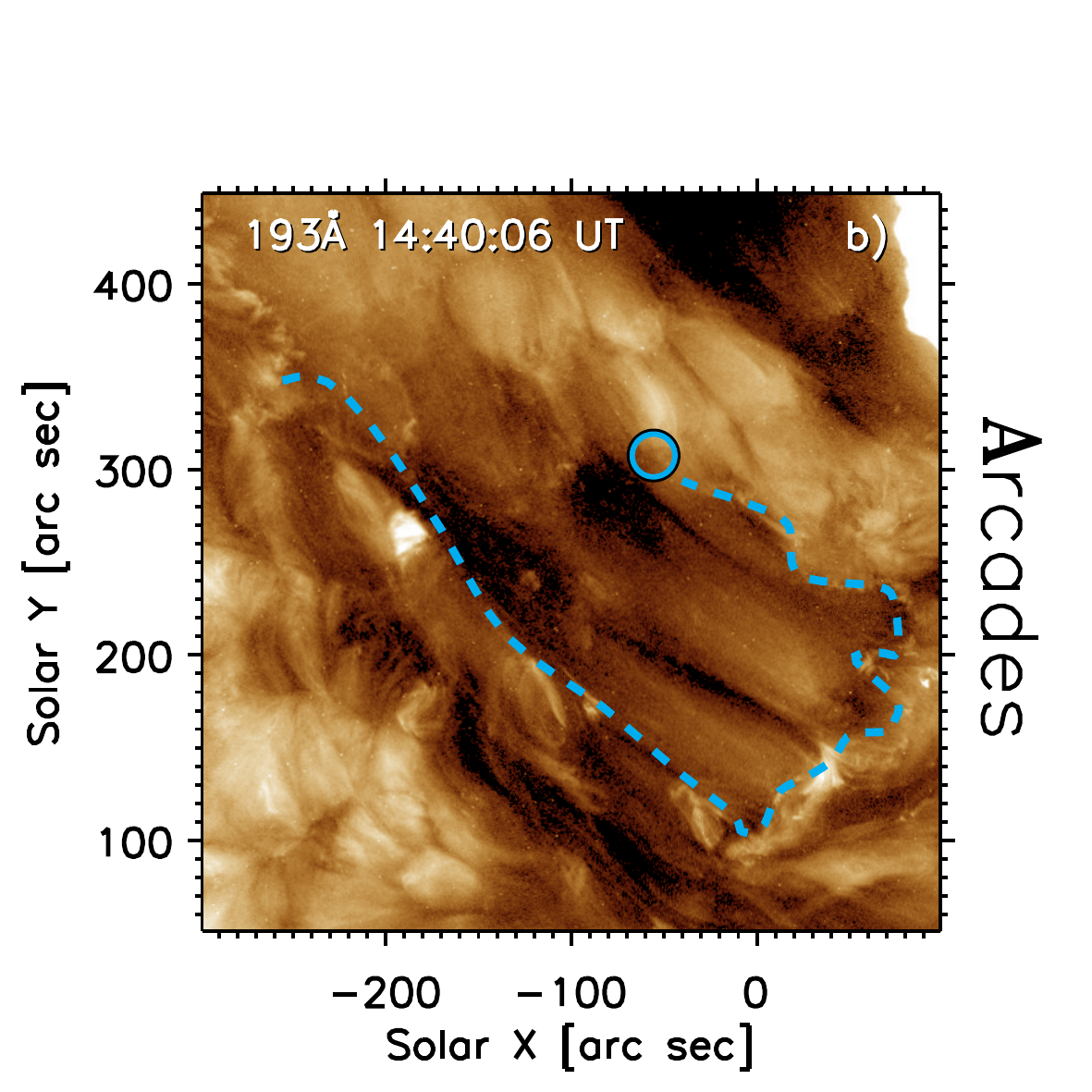}
    \includegraphics[width=4.03cm, clip,   viewport=  60 45 300 285]{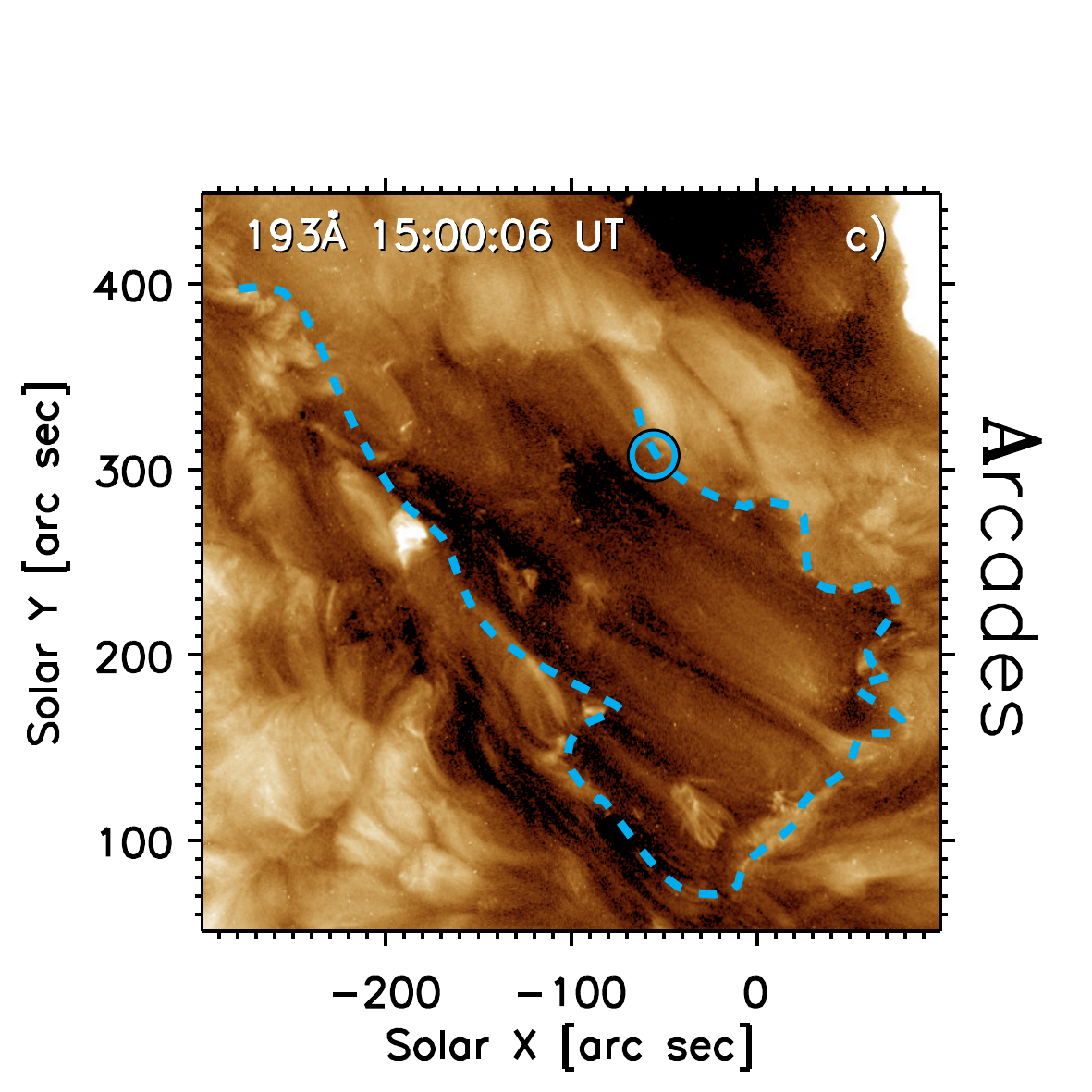}
    \includegraphics[width=4.54cm, clip,   viewport=  60 45 330 285]{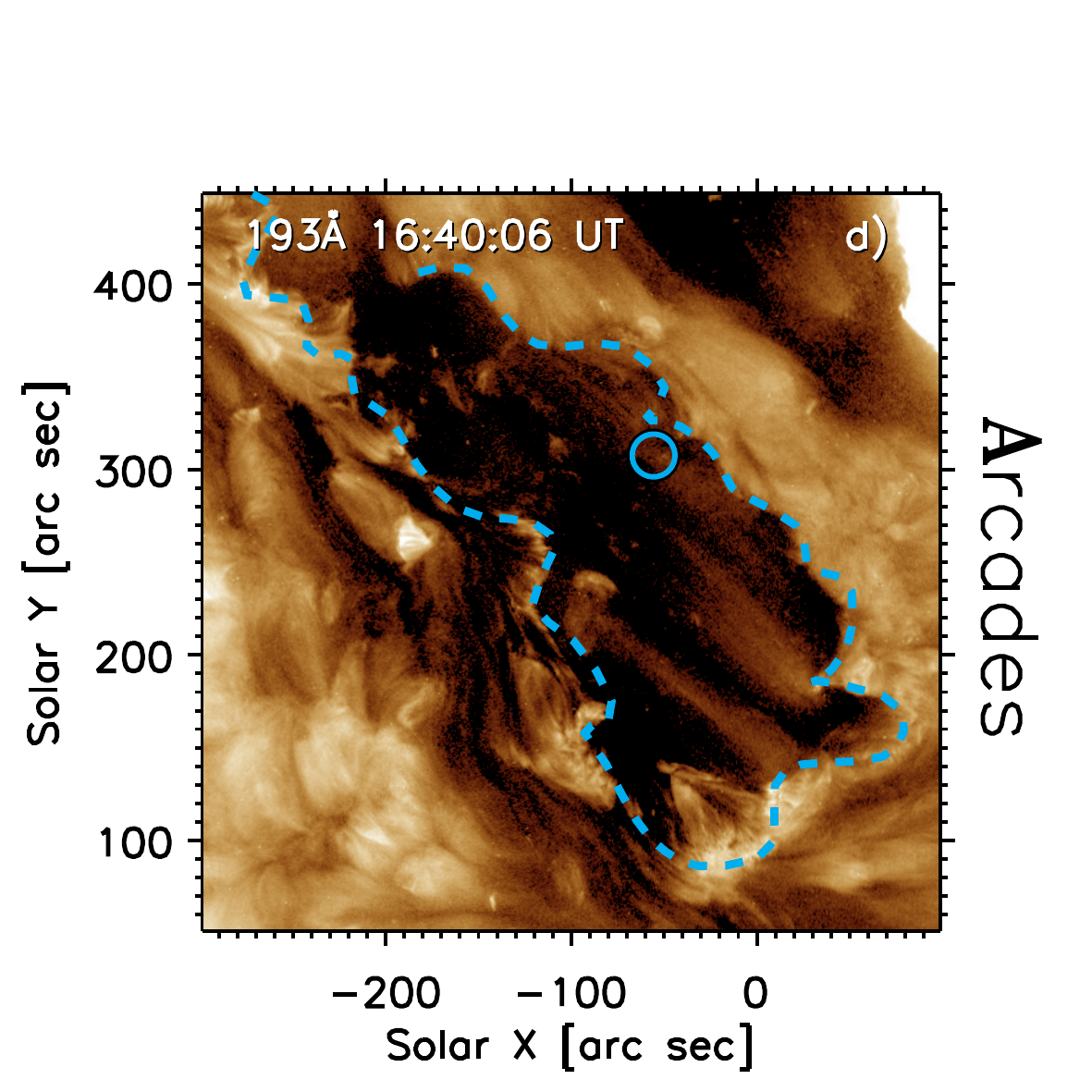}
	\\
    \includegraphics[width=5.03cm, clip,   viewport=  00 45 300 300]{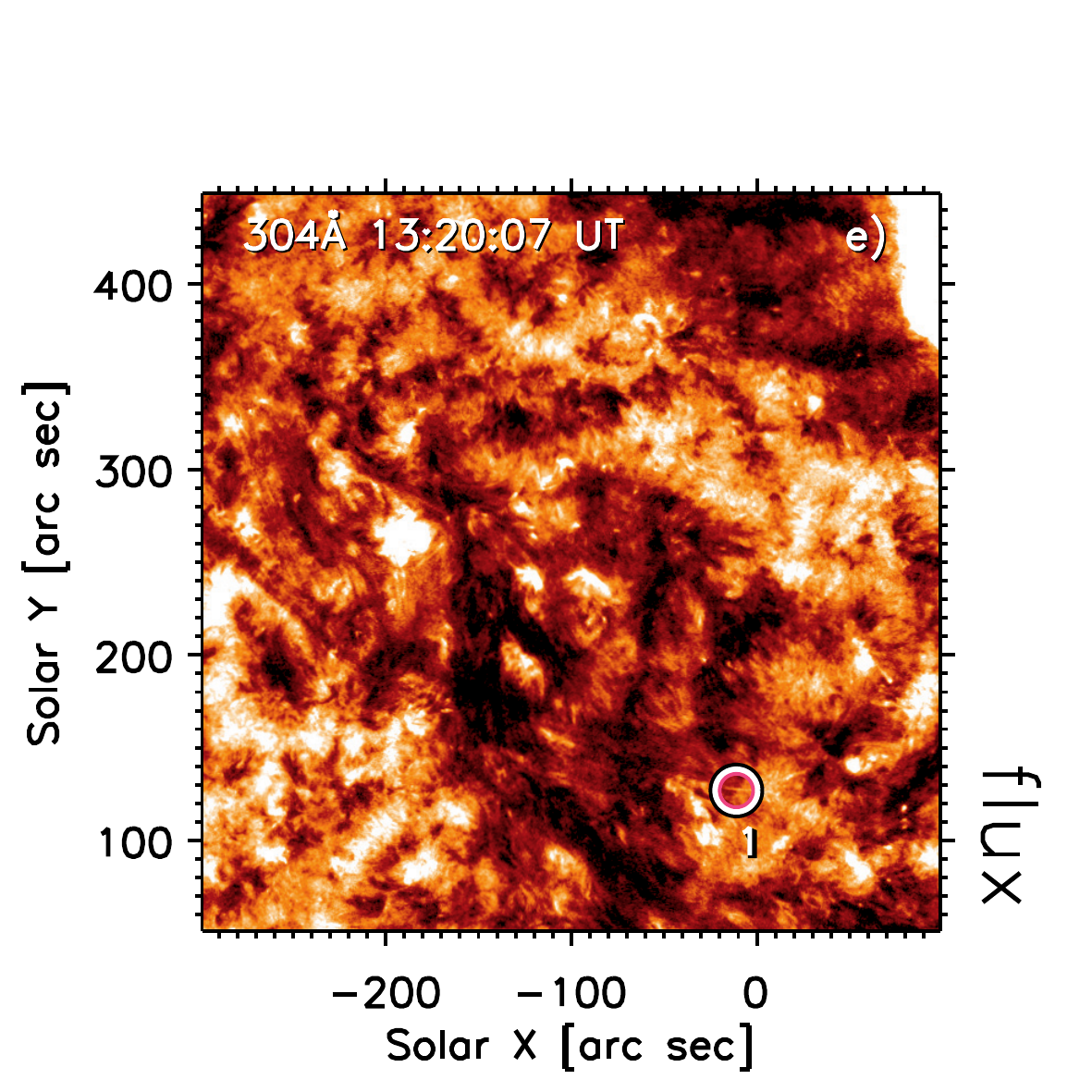}
    \includegraphics[width=4.03cm, clip,   viewport=  60 45 300 300]{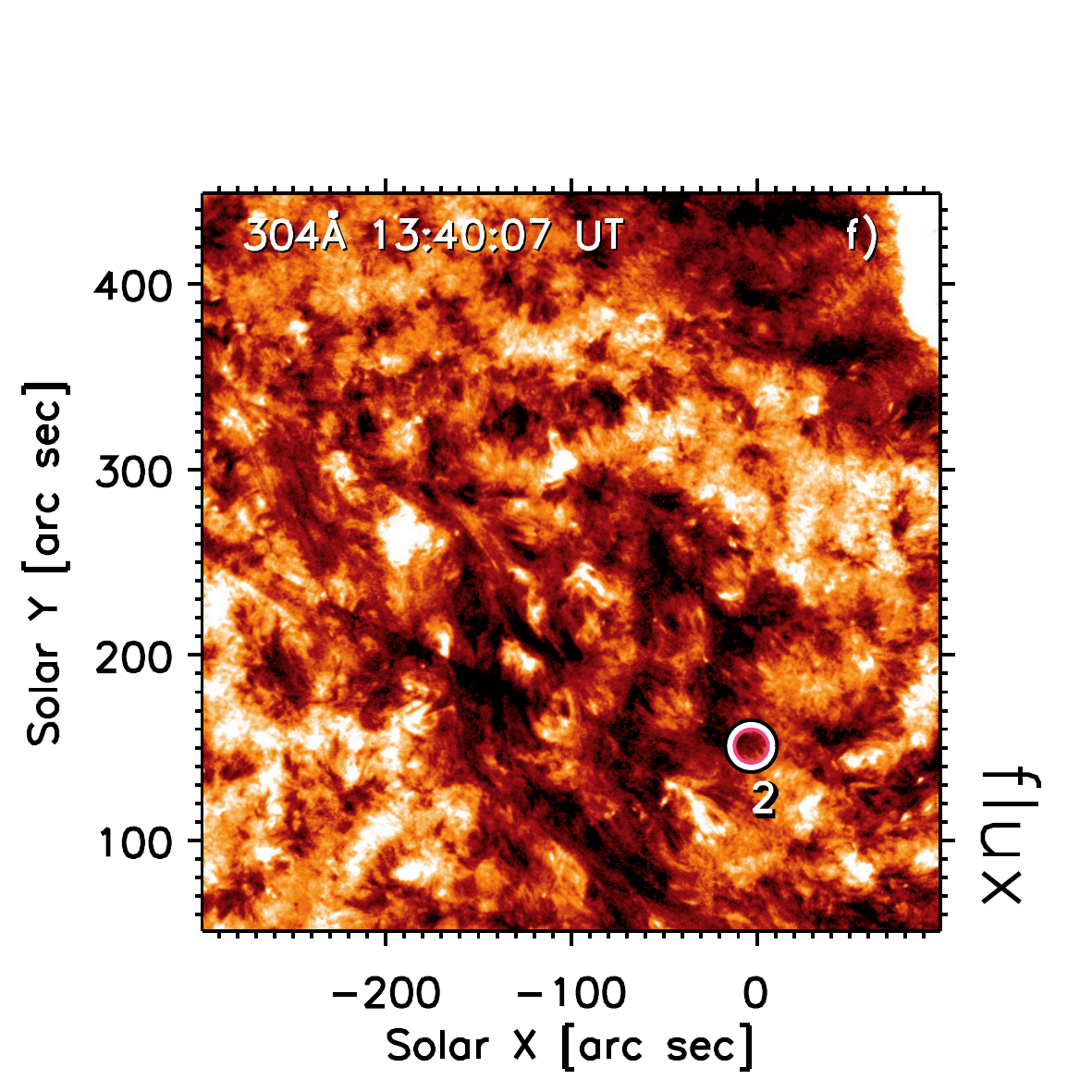}
    \includegraphics[width=4.03cm, clip,   viewport=  60 45 300 300]{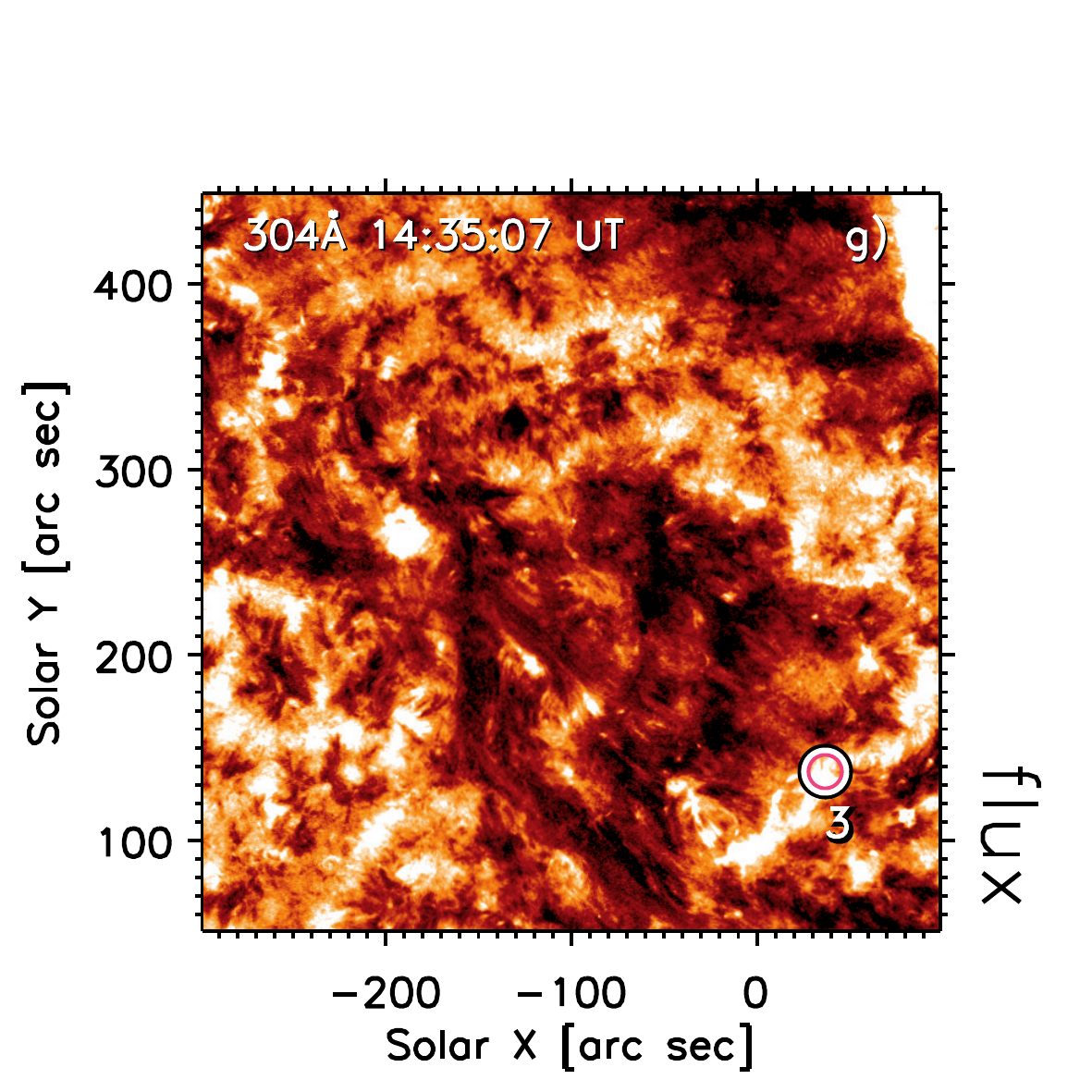}
    \includegraphics[width=4.54cm, clip,   viewport=  60 45 330 300]{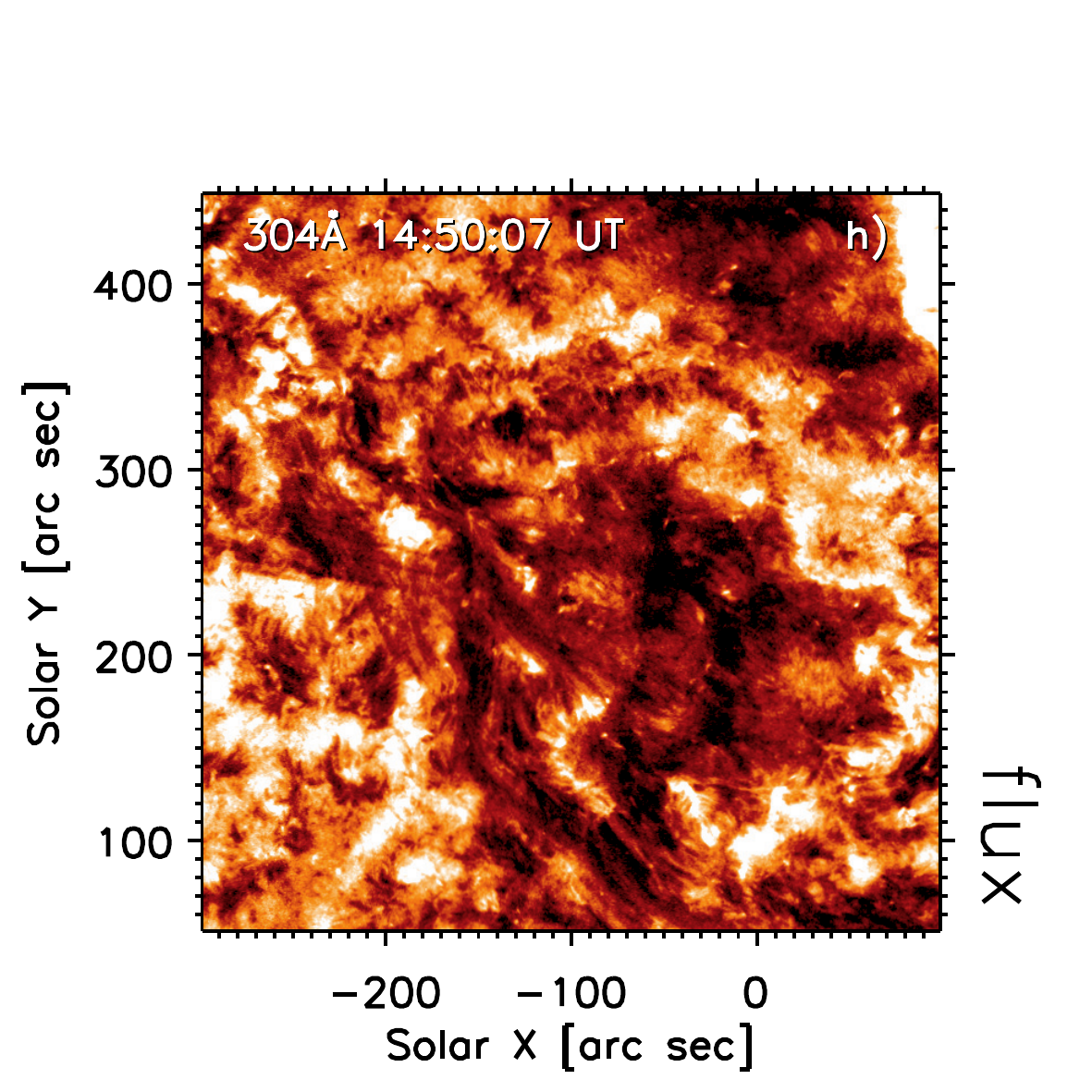}
    \\
    \includegraphics[width=5.03cm, clip,   viewport=  00 00 300 285]{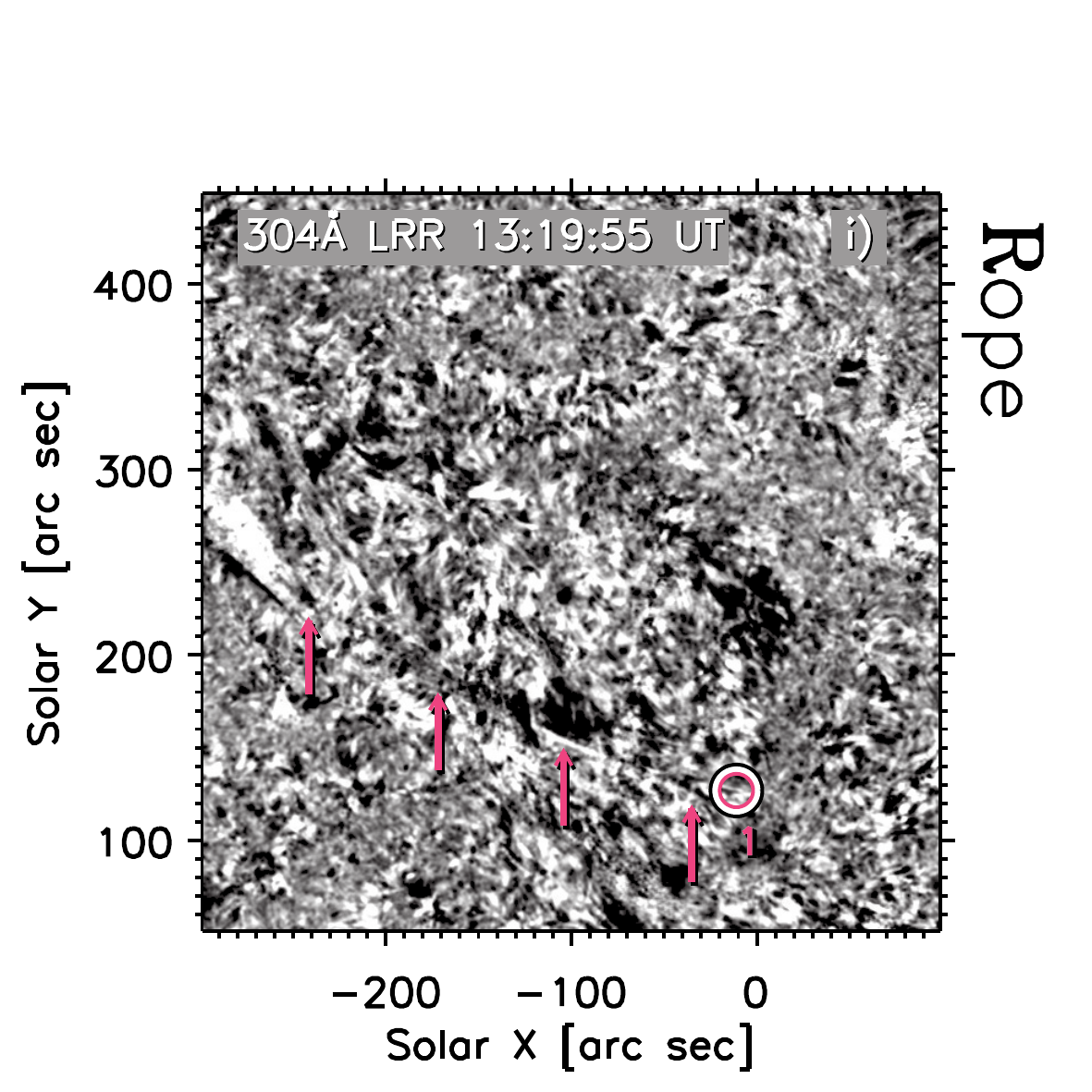}
    \includegraphics[width=4.03cm, clip,   viewport=  60 00 300 285]{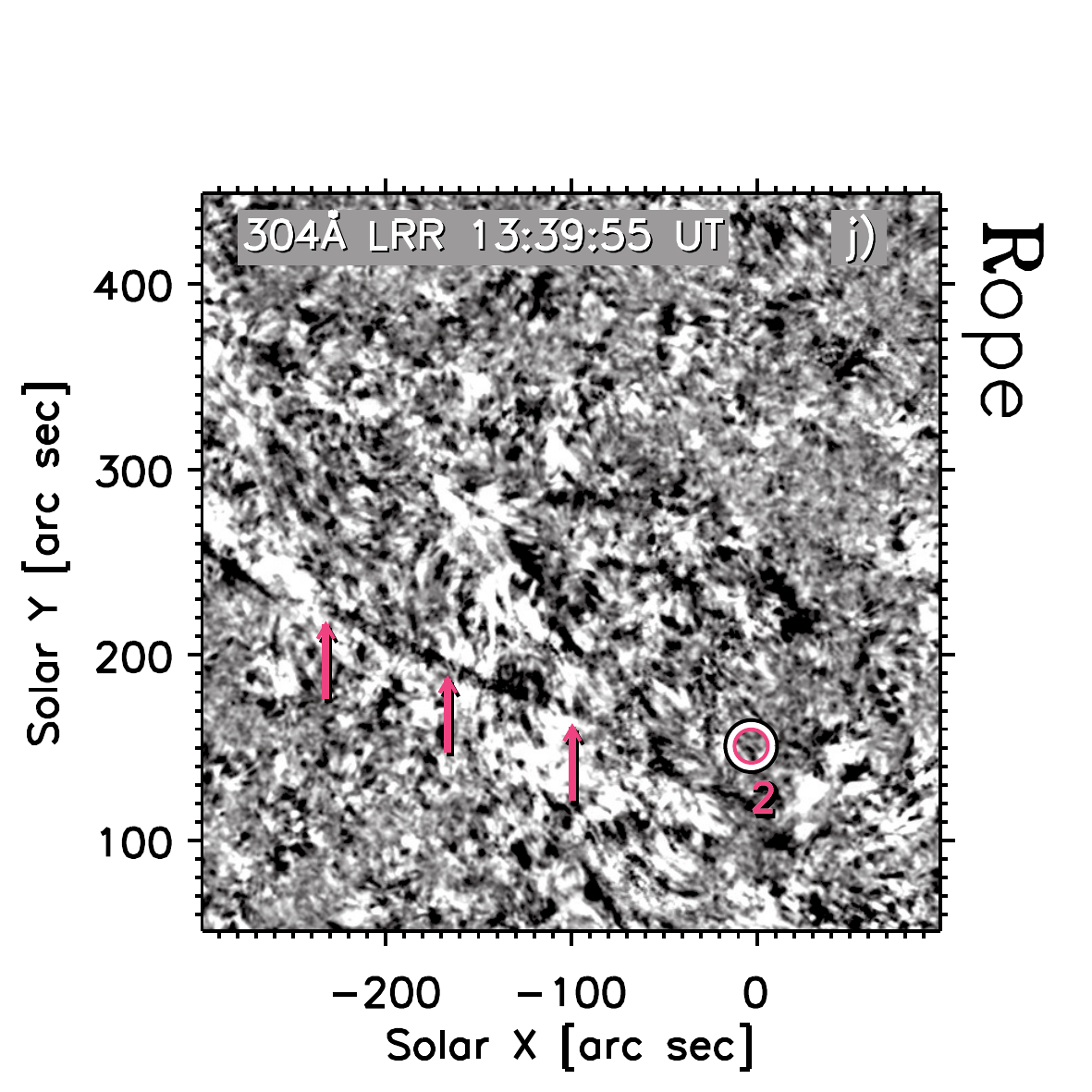}
    \includegraphics[width=4.03cm, clip,   viewport=  60 00 300 285]{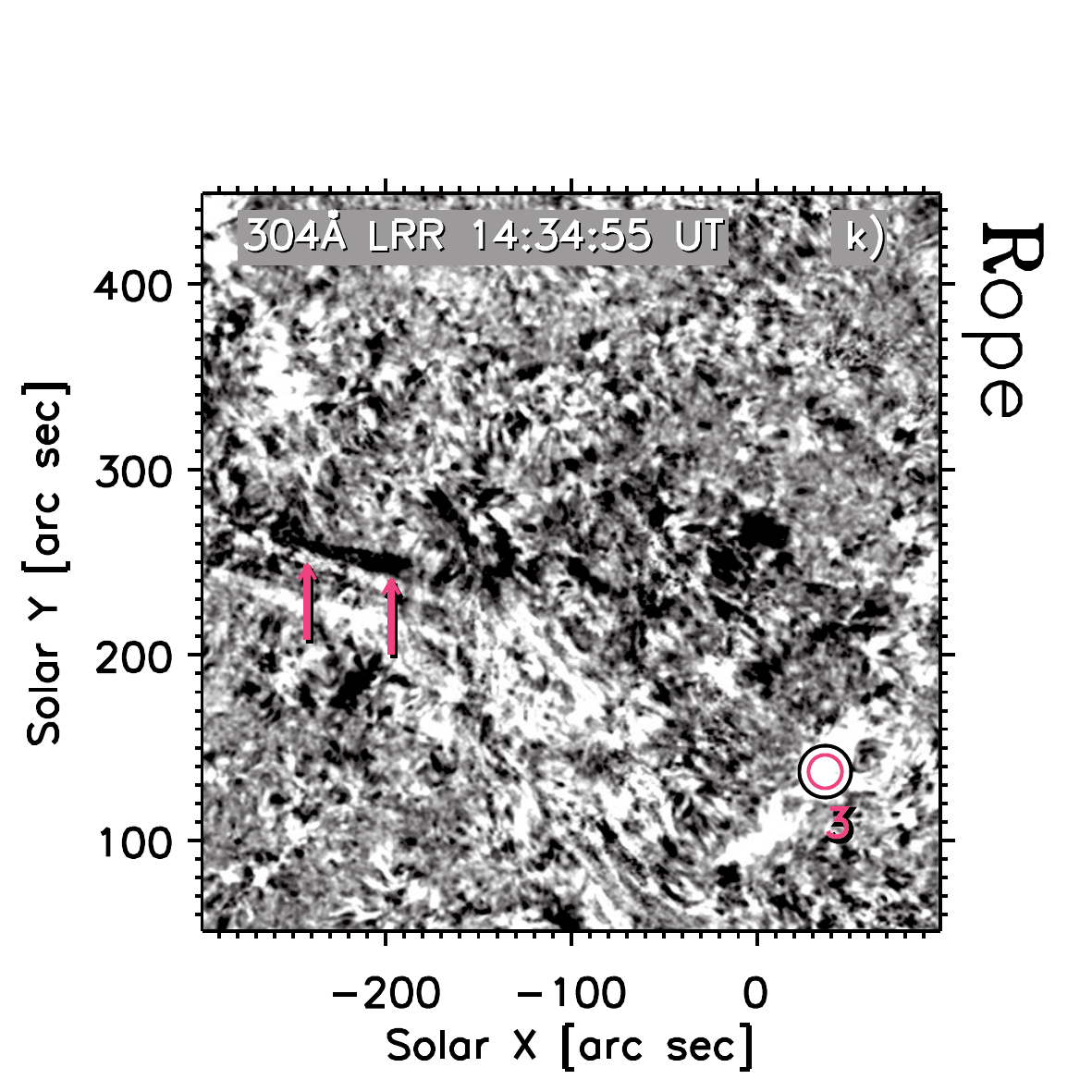}
    \includegraphics[width=4.54cm, clip,   viewport=  60 00 330 285]{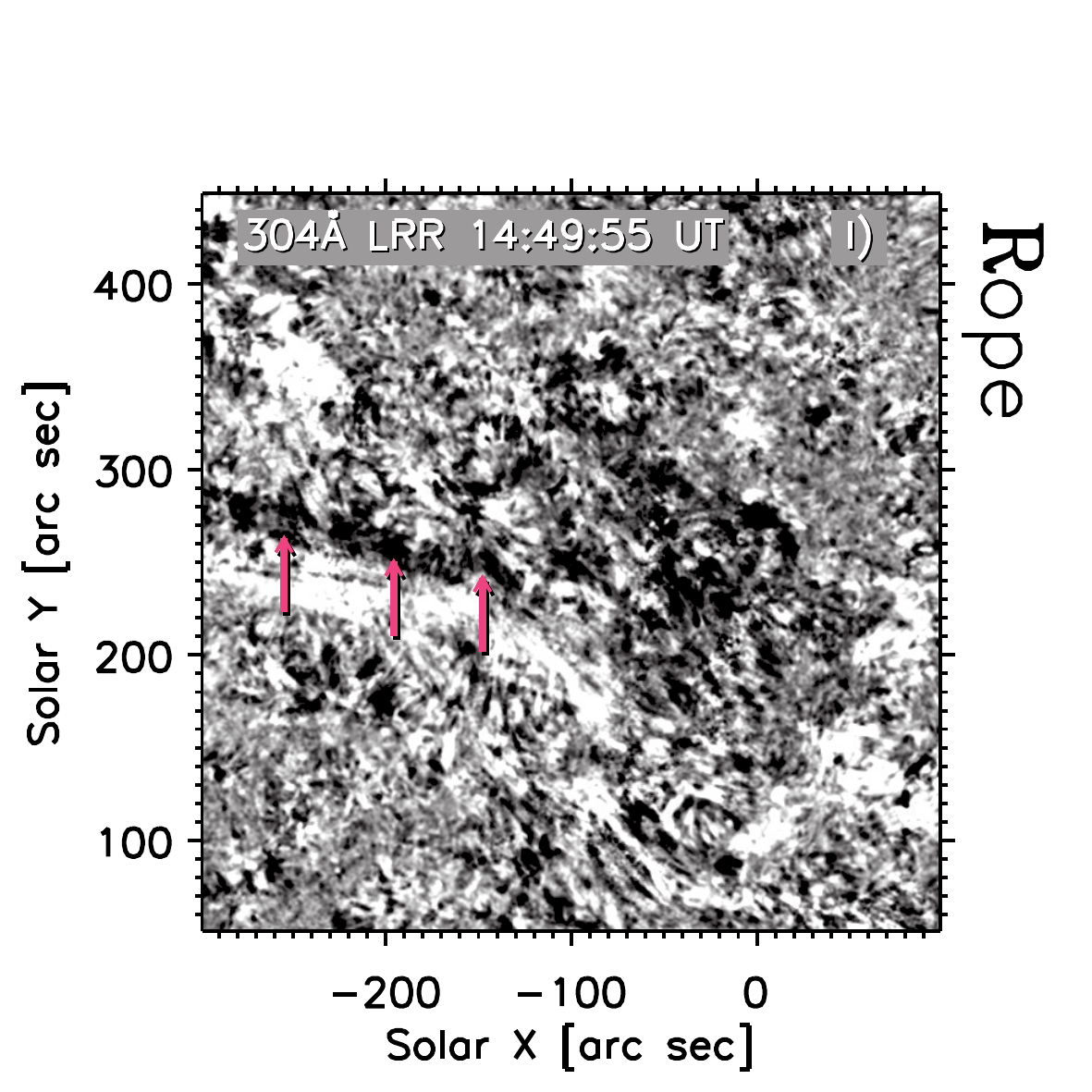}
    \\
    \includegraphics[width=9.12cm, clip, viewport=	04 0 308 215]{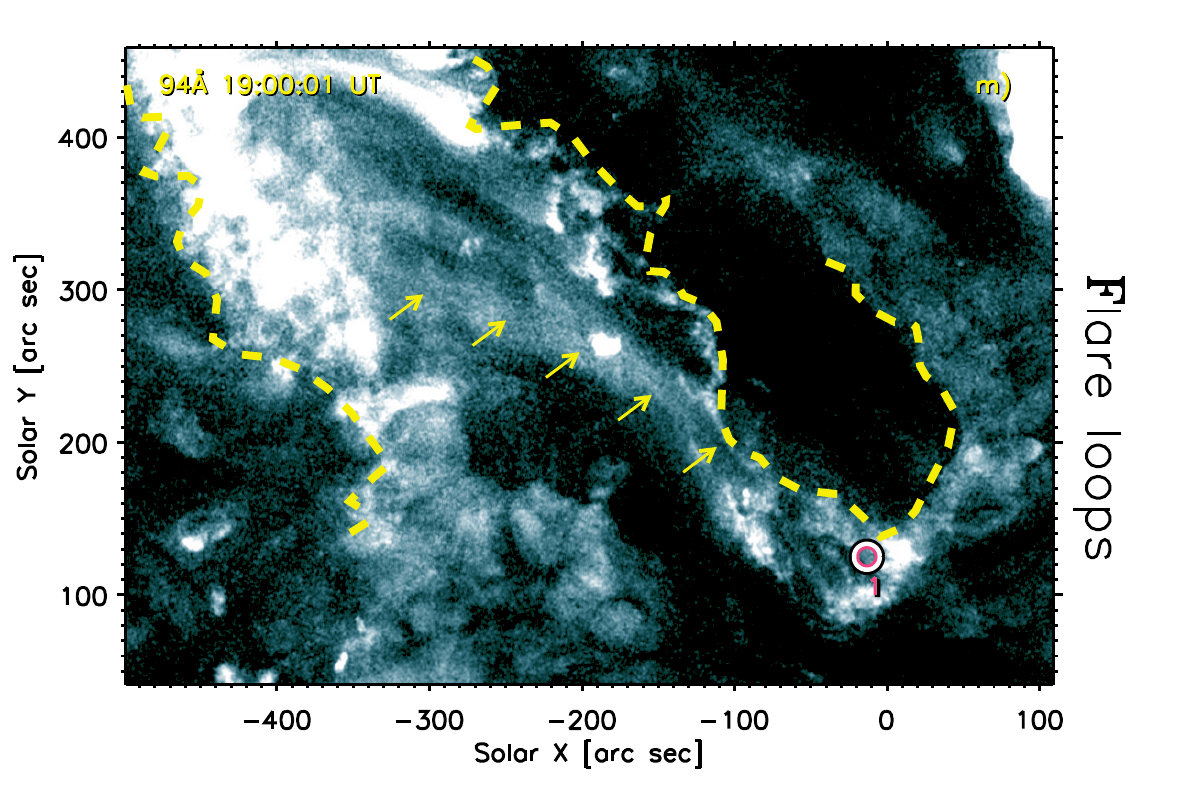}
    \includegraphics[width=8.88cm, clip, viewport=	34 0 330 215]{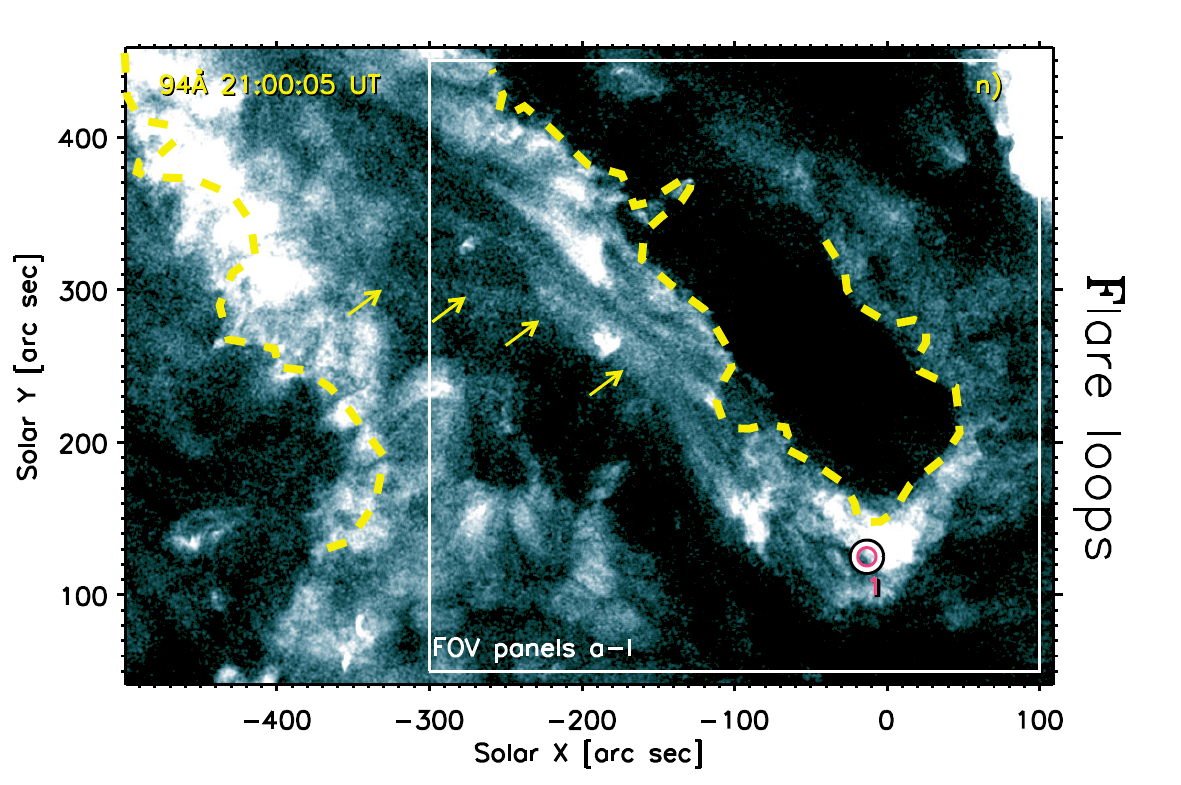}
    \caption{Individual constituents of the ar-rf reconnection. In the first row (panels (a)--(d)) we show conversion of the selected canopy (blue circle) into the erupting flux rope, as observed in the 193\,\AA~channel. The blue dashed lines mark the edge of the hook selected manually using the 304\,\AA~channel data. The evolution of the filament's threads and footpoints is presented in the second and the third row in the 304\,\AA~channel data (panels (e)--(h)), and {log-running ratio (LRR)} images (panels (i)--(l)). There, the moving filament's threads are marked using the pink arrows, while their footpoints using the white-pink circles. The hot emission as viewed in the 94\,\AA~channel is included the last row. Panels (m)--(n) show that the region where the filament was rooted before the eruption (white-pink circle) now corresponds to the footpoints of flare loops highlighted using the yellow arrows. The yellow dashed lines mark both ribbons and the hook.\\ {Animated version of the 304\,\AA~filter channel observations (panels (e)--(l)) is available in the online journal. The animation starts at 13:10 UT, ends at 15:30 UT, and its real-time duration is 35 seconds. \label{fig_arrf} }}
\end{figure*}

\section{3D Magnetic reconnection at the hook} \label{sec_arrf}

The eruption of the filament F1 was accompanied by reorganization of structures rooted in the region previously hosting its footpoints (see Section \ref{sec_hookform}). In this section we investigate phenomena associated with the formation and evolution of the hook observed in different filter channels of AIA and discuss whether these correspond to the manifestations of the ar--rf reconnection between the erupting filament and its overlying arcades, as introduced by \citet{aulanier19}. 

\subsection{Mechanism} \label{sec_arrf_mech}

The three-dimensional ar--rf reconnection describes how the arcades (`$a$') overlying the flux rope reconnect with its leg (`$r$'), into new lines composing the flux rope (`$r$') and flare loops (`$f$'). It involves the hook; during its expansion, its outer edge sweeps arcades which become a part of the erupting flux rope. They swap their connectivities with field lines composing the flux rope, swept by the inner portion of the hook. Consequently, these then turn to flare loops. The sequential change of connectivity between the arcades and the field lines composing the flux rope results in apparent drifting of flux rope footpoints along the hook.

Simultaneous observations of all constituents of the ar-rf reconnection are rare so far; as not all reconnecting structures are visible, in particular in case of dimmings. One-to-one correspondence between the observed and the predicted manifestations was reported in one event only \citep{dudik19}. Its indications were however reported in a few studies. The evolution of flare ribbons analysed by \citet{zemanova19} was found to be consistent with the ar--rf reconnection predictions. Field lines composing the erupting hot channel, were however not distinguished. Similar results were achieved by \citet{chen19}, who however reported on the drift of the flux rope footpoints based on the motion of flare kernels. \citet{lorincik19b} observed a conversion of filament threads into flare loops accompanied by a disappearance of arcades rooted near the hook. Drifting of the filament threads toward the tip of the hook was however not observed. The authors speculated that some of the post-reconnective filament field lines might have reconnected into the dimming region located in the hook and therefore were not observed. 

\subsection{Overlying arcades} \label{sec_arcades}

As mentioned in Section \ref{sec_hookform}, the leg of the filament was, before the eruption, rooted in the quiet-Sun region with multiple funnel-shaped loop footpoints. These can be divided into two groups; those hosting the outflows (funnels, Section \ref{sec_flows_dr}) and those which do not (canopies, Section \ref{sec_longterm}). Some of the canopies located to the {northwest} from the dimming region were swept by the outer edge of the forming the hook. This process is detailed in the first column of Figure \ref{fig_arrf}. Panel (a) shows the view of the region where the hook formed. Using the blue circle we marked a canopy which we selected for a further analysis. During the elongation of the hook (panel (b)), the tip of the hook converged towards this canopy. Twenty minutes later (panel (c)), after the hook elongated toward northeast, the portion of the canopy located near the edge of the hook disappeared. At 16:40 UT (panel (d)), when the hook further elongated in the same direction, the entire circle was a part of the dimming region formed in its interior. The canopy thus became a part of the erupting filament, resulting in a spatial expansion of the dimming region.

\subsection{Flux rope field lines}

The second row of Figure \ref{fig_arrf} shows the evolution of the ribbon hook as well as the threads composing the erupting filament in the 304\,\AA~filter channel data (panels (e)--(h)). The threads are difficult to track, mainly because they are dark and projected over the gradually forming dimming region. Therefore, to highlight their motion, we produced the log-running ratio (LRR) images using the same 304\,\AA~data. The time delay between the two frames producing one LRR image was chosen as 30 minutes, which was found to best highlight the moving thin threads over the evolving dimming region (panels (i)--(l)). In addition, to further enhance the visibility of the threads, the LRR images were smoothed with the \mbox{$3\times3$} boxcar, while both the original and the LRR images were averaged in 3 consecutive exposures. We also encourage the reader to review the animated version of these panels, where the motion of the filament threads is easy to discern.

In the LRR images, the filament threads are at each time highlighted using pink arrows. Their footpoints are in both the 304\,\AA~and the LRR marked using white-pink circles. At 13:20 UT (panels (e) and (i)) the threads composing the filament can be traced all the way to their footpoints located approximately at [0\arcsec, 120\arcsec] (circle 1). Later at 13:40 UT, the footpoints drifted slightly toward northwest (panels (f) and (j), circle 2). Because of the bright material falling along the threads, they were again observed to drift toward the west an hour later in the panels (g) and (k) (circle 3). Afterwards (panels (h) and (l)), the footpoints of the threads were no longer distinguished. The threads however shifted toward the north, which {is consistent with the understanding that the footpoints}, though not visible, drifted further along the hook. 

This drift of the filament threads and their footpoints is a clear manifestation of the drift of the CME legs in the same way as in \citet{dudik19}. Together with the evolution of the hook (Figure \ref{fig_hook}(a)), it confirms the presence of recently-identified 3D reconnection ar-rf geometry \citep{aulanier19}, which leads to the drifting of the flux rope footpoints in the hooked flare ribbons due to their reconnections with the surrounding solar corona.

\subsection{Flare loops}

Distribution of hot plasma was investigated using the 94\,\AA~filter channel data. There was only a small amount of hot emission associated with this event, which is supported by the flat curve of the \textit{GOES} X-ray flux which did not rise during the eruption (Figure \ref{fig_overview}(h)). To enhance the emission of flare loops and increase their signal-to-noise ratio, we averaged the data in 20 consecutive exposures and smoothed them with the \mbox{$3\times3$} boxcar. The resulting images are shown in Figure \ref{fig_arrf} (m) and (n). There, flare loops joining the negative-polarity ribbon with the positive-polarity ribbon and its hook are found, although they are still faint with average intensities usually below 1 \dn. These low intensities explain why there are no flare loops clearly identifiable in individual 94\,\AA~exposures, nor is there a discernible GOES peak.

The flare loops can first be distinguished at around 18:00 UT, approximately 3 hours after the formation of the hook, and are rooted in its elbow \citep[similarly as in][]{lorincik19b}. In panels (m) and (n) we indicated the initial positions of the filament's footpoints using the circle 1. As the time went by and the hook drifted toward {north}, these locations started to correspond to the footpoints of flare loops (highlighted using yellow arrows). This evidences a conversion of field lines composing the erupting filament into flare loops, which is in agreement with the predictions imposed for the ar-rf reconnection.

\subsection{Discussion on the presence of the three-dimensional magnetic reconnection}

Although we could not find all four structures that mutually change the connectivities during the 3D ar-rf reconnection, the present observations provide {significant} evidence that the ar-rf reconnection takes place.

First, the canopies surrounding the hook are swept by its outer edge, i.e., the $a \rightarrow r$ {(arcade to flux rope)} conversion takes place. Second, we observed the drift of the filament footpoints during the eruption. Third, the previous position of the filament footpoint turned to flare loops ($r \rightarrow f$). All this is piece-wise consistent with the predictions of the ar-rf reconnection geometry, and the latter cannot be explained by the traditional reconnection present in the 2D CSHKP model, where only arcades reconnect to become a flux rope and a flare loop. We note that so far, {the} majority of eruption events with hooked ribbons reported in literature \citep{aulanier19,zemanova19,lorincik19b} only {allowed to discern} partial constituents of the ar-rf reconnection geometry, {whilst providing valuable} evidence that field lines indeed reconnect in this newly-identified 3D reconnection geometry. The present event is no exception.

{Our observations indicate that the canopies rooted near the tip of the hook became a part of the erupting filament via the ar--rf reconnection. This means that closed-loops reconnect to become a part of the dimming region. Thus, their new magnetic field geometry is one favourable for development of outflows as shown in Section \ref{sec_outflows}. By introducing more and more field lines into the dimming region, the ar--rf reconnection led to its spatial expansion. }

{In contrast to that, the funnel-shaped loop footpoints, used for the analysis of the outflows (cut $c1$, Section \ref{sec_flows_dr}), did not enter the dimming region via the ar-rf reconnection. As the dimming region formed during the eruption, these structures were already rooted at its inner edge (Figure \ref{fig_hook}(f)--(g)) and were therefore not reconnecting. Instead, their transformation to true coronal funnel, with outflows, was driven by the opening of the field lines in the dimming region (Section \ref{sec_dilution}).}

%{Even though our observations indicate that the canopies rooted near the tip of the hook became a part of the erupting filament via the ar--rf reconnection, the funnel-shaped loop footpoints, later used for the analysis of the outflows (cut $c1$, Section \ref{sec_flows_dr}), did not. As the dimming region formed during the eruption, these structures were already rooted at its inner edge (Figure \ref{fig_hook}(f)--(g)) and were therefore not reconnecting. Instead, their transformation to true coronal funnel was possibly driven by the opening of the field lines in the dimming region (Section \ref{sec_dilution}).}

{Along this funnel, the outflows were observed for many hours irrespectively of the ar--rf reconnection occurring between $\approx$13:20 -- 14:50 UT near the tip of the hook. The funnel itself however did not reconnect up until $\approx$21:00 UT, when it was swept by the drifting hook (Figure \ref{fig_hook}(h)) which is a signature of magnetic reconnection \citep[see][]{aulanier19}. Interestingly, the fact that the funnel was swept by the hook and not the straight part of the ribbon shows that the funnel was a part of the erupting flux rope.}

\section{Summary} \label{sec_conclusions}

In this manuscript we studied an eruption of a quiescent filament from 2015 April 28th. We were analysing outflows of plasma occurring along funnels rooted in a core dimming region and ordinary coronal hole. Besides {this}, we were focused on the evolution of {a} ribbon hook encompassing the dimming region and its surrounding corona. The results of our observations can be summarized {as follows}:
 
\begin{enumerate}
\item{During the fast-rise phase of the eruption, a pair of flare ribbons formed, one of which showed a pronounced and well observed hook. After the onset of the eruption, a core dimming region started to form in the area encompassed by the hook. Its development was gradual and mainly related to fading of {structures} rooted in this region.}
\item{Along some of the funnels located in the dimming region, flow-like motions were observed in the 171\,\AA~and 193\,\AA~filter channels of AIA. Further analysis of the selected funnel revealed that the flows were directed outwards, {and} their velocities ranged between $\approx70$ and 140 km\,s$^{-1}$, while their mean velocity was 112 km\,s$^{-1}$. {The outflows were observed for more than 5 hours, until the analysed funnel faded.}}
\item{The onset of the outflows was gradual, co-temporal with the fast-rise phase of the eruption, and the formation of the hook and the dimming region within.}
\item{Outflows generating similar pattern were also observed in the neighboring coronal hole. Their velocities were, within measured uncertainty, comparable to those occurring in the dimming region.}
\item{During the eruption, the hook encompassing the dimming region was slowly drifting {toward the neighboring coronal hole. When it converged to its boundary, the hook opened and merged with it.}}
\item{The early phases of the eruption were accompanied by the drift of the threads composing the filament and their footpoints along a portion of the hook. Later, flare loops formed in regions where the filament was rooted prior to its eruption. Meanwhile, bundles of quiet-Sun canopies located in the neighbourhood of the hook were swept by its outer edge. These phenomena correspond to some of the manifestations of the ar--rf magnetic reconnection between the erupting filament and its overlying arcades.}
\item{{The ar--rf reconnection itself however does not accelerate the outflows. When the eruption started, the funnel along which the outflows were observed was already within the dimming region and it remained there during the whole period of the outflow observations. The ar--rf reconnection only acts to change the shape of the dimming region, by introducing more field lines into it.}}
\item{{Instead, our observations indicate that the sources of the outflows developed within the dimming regions along coronal funnels, not only during but also many hours after the onset of the filament eruption, thus along highly-stretched field-lines. These steady outflows are very similar to those occurring in other structures with fully-open field-lines, such as coronal holes.
}}
\end{enumerate}

The most important outcome of this study is the imaging evidence for the outflows originating in a core dimming {region. Both} the timing and the location of the outflow observations support the general view on the fundamental mechanism driving the formation of the core dimming regions, based on the depletion of plasma into the interplanetary space (Section \ref{sec_intro}). 

Of a particular interest is the similarity between the outflow velocities we measured using the time-distance diagrams and those resulting from blueshift measurements performed in various spectroscopic studies (see Section \ref{sec_intro}). Another important characteristic our results share with some of the spectroscopic studies is the location of sources of the most-intense outflows. Here, the outflows were best seen along the funnel rooted at the inner edge of the dimming region, which corresponds e.g. to the findings of \citet{veronig19} (see Figure 5 therein). 

Using the time-distance diagrams produced with the 171\,\AA~and 193\,\AA~filter channel data, we found that the properties of the outflows from the dimming region were not changing both along the funnel and in time. We therefore suggest that they were not associated with MHD waves. Since the observed dimming region corresponds to the footpoints of the filament and the outflows are initiated during the fast-rise phase of its eruption, the driver of the outflows is associated with the CME. However, the period during which they were occurring does not correspond to the $\approx20$-minute timescale of the dilution induced by the CME's expansion into the interplanetary space. 

Instead, the similarity between the properties of the outflows in the dimming region and the neighboring coronal hole points toward their origin being related to {the solar} wind. Based on our analysis we propose that, {during the eruption, structures surrounding and underlying the leg of the erupting filament turned into sources of solar wind, leading to the dimming region's resemblance of an ordinary coronal hole. What is more, these structures merged later after the eruption, with no trace of the dimming region being recognizable the day after the eruption. This outcome justifies the common interchangeability of the terms `transient coronal hole' and `dimming region'.} One of the implications of this interesting, though not unique, observable is that hooks of solar flare ribbons hosting dimming regions can be another building {block contributing to the expanded extent of coronal holes.} 

Still, our observations alone cannot provide a full picture of the mechanism accelerating the solar-wind like outflows and driving them for many hours. Perhaps the most puzzling question to be answered is whether the outflows are somehow associated with MHD waves. Even though in Section \ref{sec_interpretation} we stated that the upward-propagating brightenings are not signatures of waves, waves could still contribute to their acceleration just as the Alfv\'en waves do in the case of the solar wind \citep[see e.g.,][and references therein]{viall20}. Addressing this problem would however require observations of the distribution of $N_{\text{e}}(z)$ and $B(z)$, which is not possible with the current instrumentation. On the other hand, a significant contribution to the understanding of the nature of the dimming region outflows could be achieved by tracing the sources of {the solar} wind down to the solar atmosphere. To do so, imaging data need to be combined with measurements of element abundances carried out either \textit{in-situ}, or spectroscopically. Even though the newest spacecrafts such as the \textit{Parker Solar Probe} and the \textit{Solar Orbiter} could help resolve this involved task, distinguishing between the sources of {the solar} wind might still be difficult \citep{kilpua16}. 
\\

{Authors thank the careful referee whose comments helped us clarify this manuscript. Authors are grateful to the} providers of open-source software for online calls and meetings, which were essential for the completion of this work during the outbreak of the COVID-19 pandemic. J.L. and J.D. acknowledge the project 20-07908S of the Grant Agency of Czech Republic as well as insitutional support RVO: 67985815 from the Czech Academy of Sciences. This work was also supported by the Charles University, project GA UK 1130218. G.A. and B.S. thank the CNES and the Programme National Soleil Terre of the CNRS/INSU for financial support. L.G. is supported by contract  SP02H1701R from Lockheed Martin to SAO. AIA and HMI data are provided courtesy of NASA/SDO and the AIA and HMI science teams. 
\\

\textit{Facilities:} SDO, LASCO

\bibliographystyle{aasjournal}
\bibliography{hook_outflows}

\end{document}